\documentclass[12pt]{article}
\usepackage{graphicx}
\usepackage[usenames]{color}
\usepackage[a4paper]{geometry}

\newcommand{\spacemaker}[1]{
\setbox1=\hbox{$#1$}
\setbox2=\hbox to \wd1 {}
\box2
}

\begin{document}

\begin{titlepage}
\begin{flushright}
CALT-68-2609\\
ITEP-TH-68/06
\end{flushright}

\begin{center}
{\Large\bf $ $ \\ $ $ \\
Bihamiltonian structure of the classical superstring in 
$AdS_5\times S^5$
}\\
\bigskip\bigskip\bigskip
{\large Andrei Mikhailov}
\\
\bigskip\bigskip
{\it California Institute of Technology 452-48,
Pasadena CA 91125 \\
\bigskip
and\\
\bigskip
Institute for Theoretical and 
Experimental Physics, \\
117259, Bol. Cheremushkinskaya, 25, 
Moscow, Russia}\\

\vskip 1cm
\end{center}

\begin{abstract}
We discuss the bihamiltonian structure of the Metsaev-Tseytlin
superstring in $AdS_5\times S^5$.
We explicitly write down the 
boost-invariant symplectic structure for the superstring
in $AdS_5\times S^5$
and explain its relation to the standard (canonical) symplectic structure.
We discuss the geometrical meaning of the  boost-invariant symplectic
structure for the bosonic string.
\end{abstract}

\end{titlepage}
\section{Introduction}
The most well-known example of the AdS/CFT correspondence 
is the duality between the
Type IIB superstring in $AdS_5\times S^5$ and
the ${\cal N}=4$ supersymmetric Yang-Mills theory on
${\bf R}\times S^3$. There is a substantial evidence that both planar
${\cal N}=4$ Yang-Mills theory and the string worldsheet theory in $AdS_5\times S^5$
are quantum integrable systems 
\cite{Tseytlin:2003ii,Zarembo:2004hp,Beisert:2004ry,Plefka:2005bk}. 
However at this time there is no satisfactory treatment of the string worldsheet
theory as a quantum integrable system. This is partly because of the curved
configuration space, but the main reason is perhaps the conformal
invariance. The technique for dealing with the conformally invariant
integrable systems has not been very well developed. And if we gauge fix the
conformal invariance by choosing a special set of worldsheet coordinates, such as
the light-cone gauge, then we also loose the relativistic invariance.

It turns out that the classical string in $AdS_5\times S^5$ is closely related
to some other integrable systems which do have a relativistic symmetry. 
A toy example is the relation between the string on ${\bf R}\times S^2$ and
the sine-Gordon model. The sine-Gordon model is a two-dimensional integrable relativistic
field theory. On the level of classical equations of motion the two models are
essentially equivalent \cite{Pohlmeyer:1975nb}. 
The boost symmetries of the sine-Gordon model correspond
to some hidden symmetries of the classical string on ${\bf R}\times S^2$
(to be more precise, these hidden symmetries act on a string modulo the
global rotations of $S^2$).
The Poisson structure of the classical string is not 
invariant under these boosts.
But as a classical integrable system, string on ${\bf R}\times S^2$ has
an infinite family of Poisson brackets, which are in some sense mutually
compatible. One of these non-standard Poisson brackets is 
boost-invariant, and in fact coincides with the sine-Gordon symplectic
structure. 

{\bf What we will do.}
In this paper we will explicitly write down the 
boost-invariant symplectic structure for the superstring
in $AdS_5\times S^5$
and explain its relation to the standard (canonical) symplectic structure.
We first derive the canonical symplectic structure from the string
worldsheet action. We then construct some one-parameter group
of symmetries of the classical
solutions, which is a generalization of the relativistic boosts
of the sine-Gordon model. We find that the canonical Poisson structure
is not invariant under this one-parameter group of symmetries. However,
the canonical Poisson bracket can be written as a sum of three terms:
\begin{equation}\label{InhomogeneousPoissonBrackets}
	\{F,G\}^{can}=
	\{F,G\}^{[-2]}+\{F,G\}^{[0]}+\{F,G\}^{[2]}
\end{equation}
where the middle term $\{F,G\}^{[0]}$ is boost-invariant, and the
terms $\{,\}^{[-2]}$ and $\{,\}^{[2]}$ have scaling degrees
$-2$ and $2$. This means that if $B_{\lambda}$ is the boost transformation
with the parameter $\lambda$, then 
\begin{equation}\label{ScalingDegree}
	\{B_{\lambda}^*F,B_{\lambda}^*G\}^{[\pm 2]}=\lambda^{\pm 2}
	\{F,G\}^{[\pm 2]}
\end{equation}
We explicitly verify that the bracket $\{,\}^{[0]}$ satisfies the 
Jacobi identities. The Jacobi identity for the canonical bracket
(\ref{InhomogeneousPoissonBrackets}) follows from its construction
as a canonical Poisson bracket.
Then Eqs. (\ref{InhomogeneousPoissonBrackets}) and (\ref{ScalingDegree})
immediately imply that $\{,\}^{[2]}$ and $\{,\}^{[-2]}$ also satisfy the Jacobi
identities, and moreover are compatible with $\{,\}^{[0]}$. 
We then show that $\{,\}^{[-2]}$ can in fact be expressed in
terms of $\{,\}^{[0]}$ and $\{,\}^{[2]}$ 
by Eq. (\ref{RelationBetweenBrackets}).
Finally, we give a geometric
interpretation of the boost-invariant bracket $\{,\}^{[0]}$ in the purely
bosonic case, Eq. (\ref{BoostInvariantGeometric}).

{\bf Earlier work.} Essentially the same results, in the bosonic
sector, were previously obtained in 
\cite{Doliwa:1994bk,MR1701825,MR1102831,MR1894465,MR2058803,MR2068787,MR2139528,MR2180896,Anco:2005a,Anco:2005b}, but from a different perspective. 
The main difference of our approach is that we start from the 
relativistic string and derive all the Poisson structures
from the canonical Poisson structure of the string worldsheet action. 
The case of string on ${\bf R}\times S^2$ was considered
in our previous paper \cite{Mikhailov:2005sy}.
The thorough analysis of the equal-time Poisson brackets in the bosonic
sector was presented in \cite{Dorey:2006mx}; see also
\cite{Das:2004hy,Das:2005hp} for an earlier work. In our paper
we concentrate on the light-cone Poisson brackets. 
Also we use the currents of the generalized sine-Gordon model,
in order to make the  action of the boosts more  transparent.
The light-cone approach to the nonlinear sigma-model was previously
used to study Poisson brackets in \cite{DasLightfront}.
Poisson brackets in the pure spinor model were studied in 
\cite{Bianchi:2006im}.

{\bf Plan of the paper.}
In Section \ref{sec:PoissonStructureGeneral} we will review
the general definition
of the symplectic structure. 
In Section \ref{sec:NLSM} we will review the nonlinear
sigma-model, its relation to the generalized sine-Gordon model, and the hidden
relativistic symmetry. Section \ref{sec:Superstring} is a review of the
classical superstring in $AdS_5\times S^5$ and its canonical Poisson structure.
In Section \ref{sec:Bihamiltonian} we derive the bihamiltonian structure
of the classical superstring and discuss its properties.
In Section \ref{sec:Geometrical} give a geometrical interpretation of the
boost-invariant Poisson bracket. 

\section{General facts about Poisson brackets and symplectic structure}
\label{sec:PoissonStructureGeneral}
Poisson brackets are very important in the classical mechanics, in particular
because they are the classical analogue of the quantum mechanical commutators.
Poisson brackets are closely related to the symplectic form, and in fact
we will use both concepts simultaneously. 

\subsection{Symplectic form}
The symplectic form is computed directly from the classical action, in the
following way.
Suppose that we have a classical field theory with the action
\[
S=\int d\tau^+d\tau^- {\cal L}[\phi]
\]
We usually compute the action over the infinite space-time, but let us suppose
that we decided to compute the action in a
finite region of $\tau^+,\tau^-$. Let us take $\phi=\phi_{cl}$ a classical solution.

\begin{centering}
\begin{minipage}[t]{0.40\linewidth}
\raisebox{-1.6in}{
\includegraphics[width=1.6in]{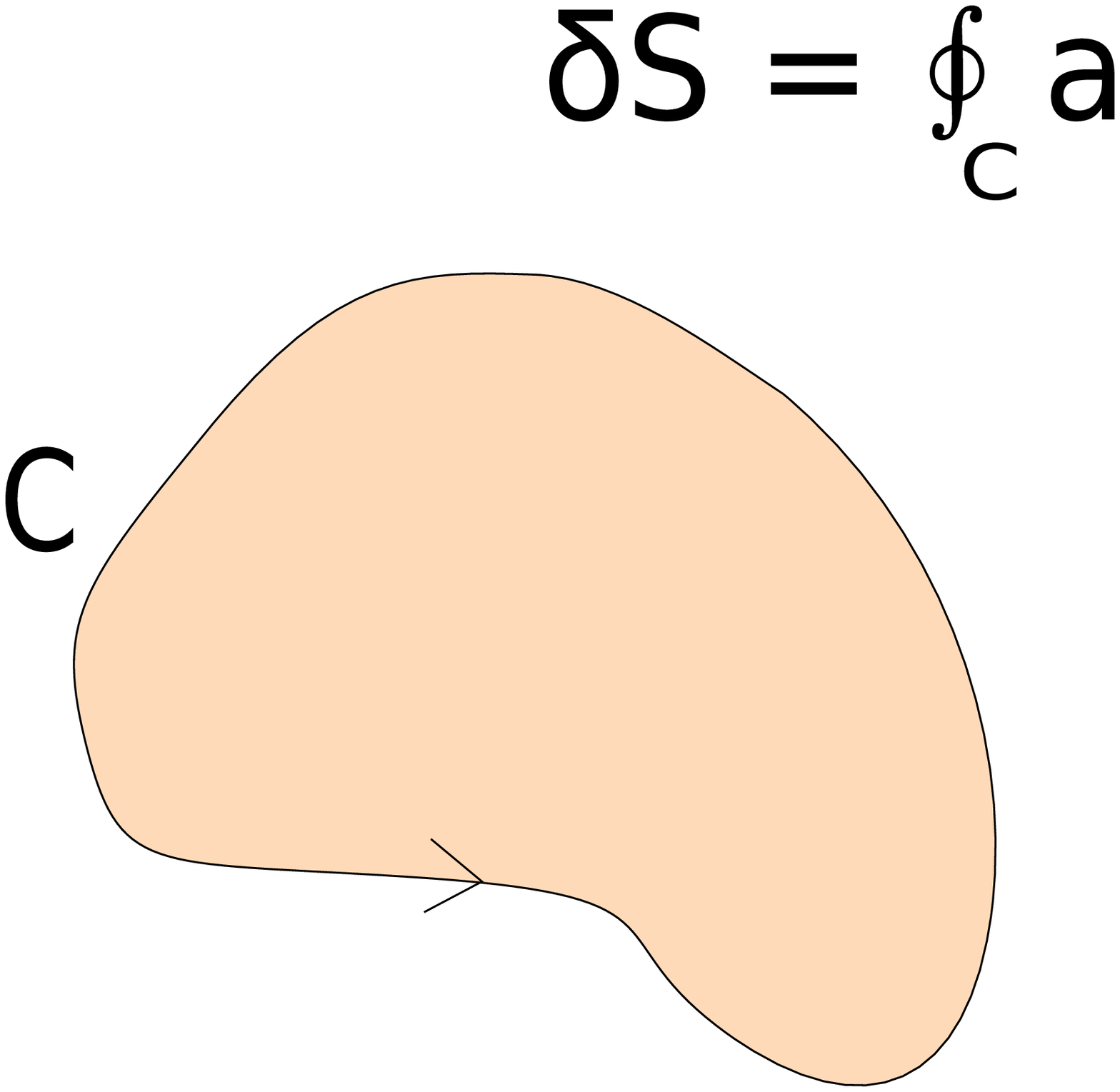}}
\end{minipage}\hfill
\begin{minipage}[t]{0.60\linewidth}
Suppose that we change $\phi_{cl}$ by a small 
amount $\delta \phi$. We will not require that $\phi_{cl}+\delta\phi$ is again a classical
solution, in other words that $\delta\phi$ is ``on-shell". 
But because $\phi_{cl}$ is an extremum of the action, for any $\delta\phi$ we will get: 
$
\delta S=\oint_C a
$
where $a$ is some 1-form on the worldsheet. 
If we restrict $\delta\phi$ to be a classical variation $\delta\phi_{cl}$ of the 
classical solution, then we get $\delta S$ a 1-form on the phase space of the system,
which is the space of all classical solutions.  Since the expression for $a$ contains both
$\delta \phi$ and $d\tau$ we can say that $a$ is ``a form of the type $(d\tau) (\delta \phi)$".
\end{minipage}
\end{centering}

\vspace{10pt}
\noindent
Let us restrict $a$ to the tangent space to the space of classical 
solutions and consider $\omega=\delta a$. This is a form of the type 
$(d\tau)(\delta\phi_{cl})^2$, where $\delta\phi_{cl}$ is now on-shell.
Notice that $\omega$ is defined unambiguously, modulo adding
a $d$-exact form. This is because $a$ is defined unambiguously
modulo a $d$-exact form, because we defined $a$ as a restriction
of the unambiguously defined expression with $\delta\phi$ off-shell.
Notice that $\omega$ is $d$-closed, since $\omega=\delta a$ and 
$da$ is $\delta$ of the action density. 
To define the symplectic form we
 consider the theory on a cylinder:
\vspace{10pt}

\begin{centering}
\hfill
\includegraphics[width=2in]{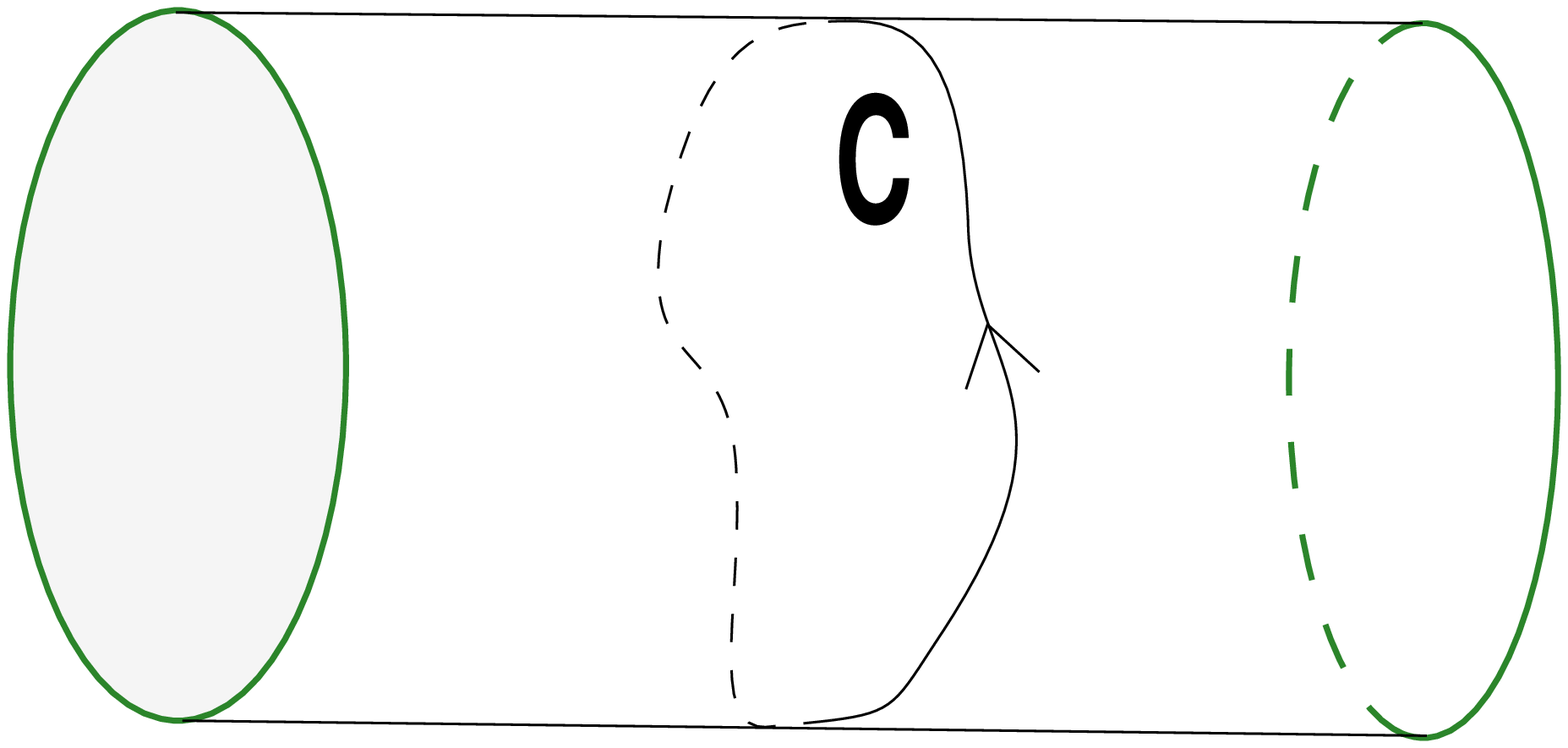}
\hfill
\end{centering}

\noindent
The symplectic form is by definition:
\[
\Omega=\oint_C \omega
\]
      This is a closed 2-form on the phase space. 
It is also sometimes useful to consider the ``symplectic potential'' which
	is defined as $\delta^{-1}$ of the symplectic form: 
\begin{equation}\label{DefSymplecticPotential}
\alpha = \oint_C a
\end{equation}
We have
$
\delta\alpha=\Omega
$
but we have to remember that $\alpha$ depends on the choice of the contour $C$.

For example, consider the particle with the action 
$\int dt \left({\dot{x}^2\over 2}-V(x)\right)$.
The phase space can be defined as the space of solutions of the classical equations
of motion $\ddot{x}=-V'$. The phase space is equipped with the "local" 2-form:
\[
\Omega= \delta \dot{x}(t_0)\wedge \delta x(t_0)
\]
which is local in a sense that it requires evaluation of the classical field
variables $x$ and $\dot{x}$ at one point in time $t=t_0$ and the result of the
calculation of $\Omega$  does not depend on the choice of $t_0$. 
In the $d+1$-dimensional classical field theory, with the kinetic
term $\int dx^0\cdots dx^d \partial_{\mu}\Phi \partial^{\mu}\Phi$ we have
\[
\Omega=\int_C \delta\Phi\wedge *d\delta \Phi
\]
where $C$ is a $d$-dimensional contour, and $\Omega$ does
not depend on the choice of this contour.

\subsection{Poisson bracket}
The Poisson bracket $\{\cdot , \cdot\}$ is defined by a bivector $\theta^{ij}$
which is the inverse of $\Omega_{ij}$:
\[
\theta^{ik}\Omega_{kj}=\delta^i_j
\]
The Poisson bracket of two functions $F$ and $G$ on the phase space is 
defined by the formula:
\[
\theta=\theta^{ij}{\partial\over\partial \Phi^{i}}\wedge 
{\partial\over\partial \Phi^{j}}
\]
The fact that $\Omega$ is closed translates to the {\em Jacobi identity}
of the Poisson bracket:
\begin{equation}\label{JacobiIdentity}
	\{F,\{G,H\}\}+
	\{G,\{H,F\}\}+
	\{H,\{F,G\}\}=0
\end{equation}
In terms of $\theta$ the Jacobi identity is some bilinear equation:
\begin{equation}\label{SchoutenBracketZero}
	[\![ \theta,\theta ]\!]=0
\end{equation}
The operation $[\![,]\!]$ on bivectors is the so-called {\em Schouten bracket}.
This is the most natural extension of the Lie bracket (the commutator of the
vector fields) to the bivectors. 

\subsection{The lightcone approach to the description of the
symplectic structure}
\label{sec:LightconeApproach}
As we have explained, the symplectic form on the phase space of classical
solutions of the two-dimensional field theory
is given by $\oint_C \omega$ where $\omega$ is something like
$\delta\varphi\wedge *d\delta\varphi$, and the integral does
not depend on the choice of the contour $C$ because $\omega$ is
$d$-closed.  

\begin{centering}
\begin{minipage}[t]{0.6\linewidth}
On the string worldsheet, through every point pass two
light-like curves (the light cone). We will call the two light-like curves
$C^+$ and $C^-$.
These curves are called characteristics.
\end{minipage}
\begin{minipage}[t]{0.4\linewidth}
\begin{centering}
\hfill
	\raisebox{-0.6in}{
	\includegraphics[width=0.7in]{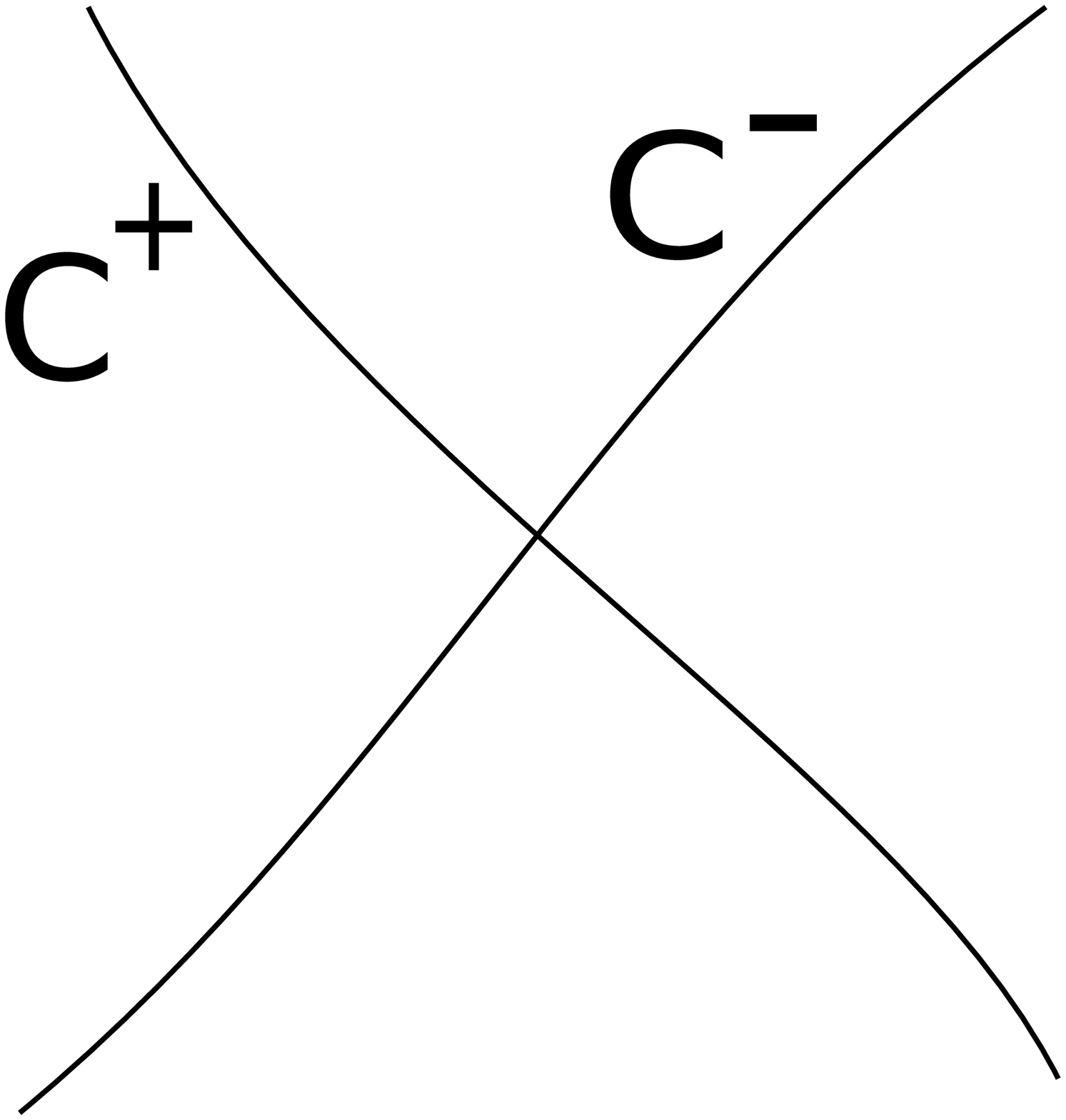}
	}
\hfill
\end{centering}
\end{minipage}
\end{centering}

\noindent
Consider an infinite string worldsheet, and choose the saw-like contour $C$ 
interpolating between two spacial infinities, consisting of the
light-like pieces:

\begin{centering}
\hfill
	\includegraphics[width=3in]{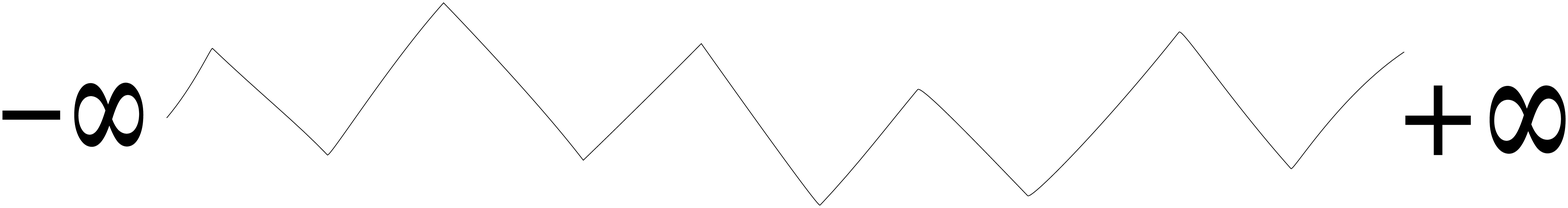}
\hfill
\end{centering}

\noindent
For massive theories, there are excitations which are spacially localized
(like breathers of the sine-Gordon model). 

\begin{centering}
\begin{minipage}[t]{0.6\linewidth}
If the tooth of the saw is
sufficiently large, we can imagine that the intersection of the breather
with the contour fits essentially (modulo the exponentially
decreasing tails) into one light-like piece.
This suggests that the classical solutions rapidly decreasing at
infinity can be
described in terms of their intersection with the characteristic $C^+$.
\end{minipage}
\begin{minipage}[t]{0.4\linewidth}
\begin{centering}
	\hfill
\raisebox{-1in}{
	\includegraphics[width=1.4in]{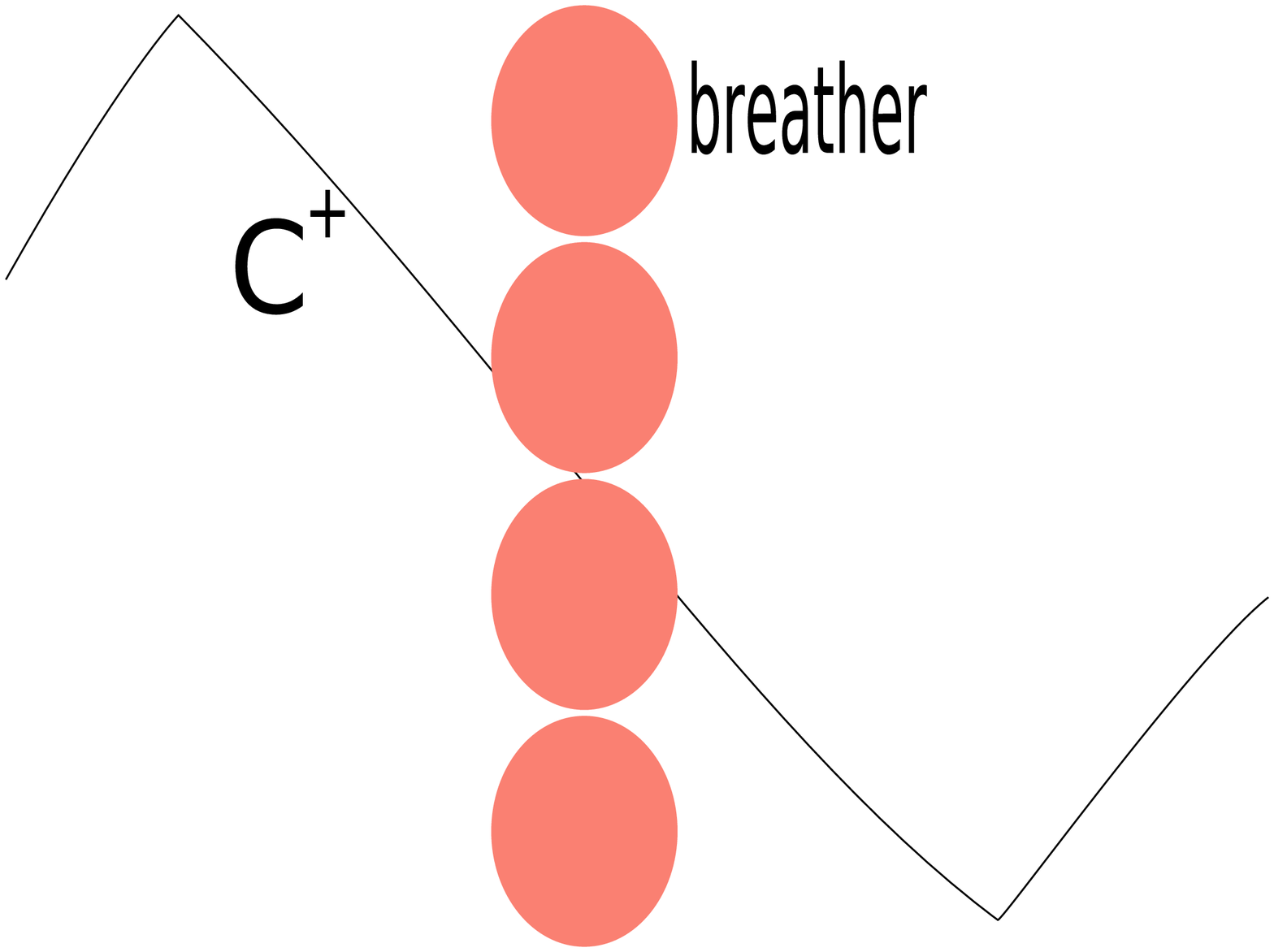}
}
	\hfill
\end{centering}
\end{minipage}
\end{centering}

\noindent
For example, consider the characteristic given by the
equation $\tau^-=0$. A rapidly decreasing solution $\varphi(\tau^+,\tau^-)$
is characterized by a function $\varphi(\tau^+)=\varphi(\tau^+,0)$.
The symplectic structure is given by the integral of $\omega$ over
the characteristic $C^+$:
\begin{equation}
	\Omega=\int_{-\infty}^{+\infty} d\tau^+
	\delta \varphi(\tau^+)\partial_+\delta\varphi(\tau^+)
\end{equation}
We will call this the ``light-cone symplectic structure''.
There are many subtleties with this approach to describing the
symplectic structure (there are important solutions which are not 
rapidly decreasing at the spacial infinity, see for example
\cite{Bak:1999iq,Callan:1999ki,Mikhailov:2003er}; 
if $\varphi(\tau^+,0)$ goes to zero
when $\tau^+\to\pm\infty$, what would guarantee that this is
true also at $\tau^-=\mbox{const}\neq 0$ ?).
In this paper we will neglect these subtleties.

\subsection{Compatibility of Poisson brackets}
\label{sec:Compatibility}
Two Poisson brackets $\{,\}_{1}$ and $\{,\}_{2}$ are called {\em compatible}
if their sum $\{,\}_{1}+\{,\}_{2}$ satisfies the Jacobi identity. 
Integrable systems usually have infinitely many compatible Poisson brackets. 
Actually, it is enough
to have two, and then it is possible
to generate an infinite family.
Indeed, suppose that the Poisson bivectors $\theta_{1}$ and $\theta_{2}$ 
give two compatible Poisson brackets. This means that $ [\![ \theta_{1},\theta_{1} ]\!]=0$
and $[\![ \theta_{2},\theta_{2} ]\!]=0$ and $[\![ \theta_{1},\theta_{2} ]\!]=0$.
Therefore for an arbitrary parameter $t$ the bivector $\theta_{1}+t\theta_{2}$
also satisfies (\ref{SchoutenBracketZero}) and therefore the corresponding
2-form $(\theta_{1}+t\theta_{2})^{-1}$ is closed. Given that $(\theta_{1})^{-1}$
is closed, this implies at small $t$ that $\theta_{1}^{-1}\theta_{2}\theta_{1}^{-1}$
is also closed. This means that $\theta_1\theta_2^{-1}\theta_1$ is again a Poisson 
bracket. 

The expression $N=\theta_1\theta_2^{-1}=\theta_1\Omega_2$ is an operator
acting on the sections of the tangent bundle to the phase 
space; in other words, this is a section of $TM\otimes T^*M$.
Operator $N$ has a special name ``recursion operator''. The compatibility of the 
Poisson brackets $\theta_1$ and $\theta_2$ implies vanishing of the Nijenhuis torsion
of $N$, see for example Section 2 of \cite{Falqui} for a concise review.

In this paper the compatible Hamiltonian structures will appear in the following
way. 
We will have some vector field $V$ on the phase space (the infinitesimal boost), 
and the canonical
Poisson bracket will be a sum of three pieces $\theta^{can}=\theta^{[2]}+\theta^{[0]}
+\theta^{[-2]}$, and the action of $V$ on $\theta^{can}$ (the Lie derivative) will be:
\begin{equation}
	V.\theta^{can}=	V.\left(\theta^{[2]}+\theta^{[0]}+\theta^{[-2]}\right)=
	\theta^{[2]}-\theta^{[-2]}
\end{equation}
Because of the geometrical naturalness of the Lie derivative and the
Schouten bracket we have:
\[
V.[\![\theta_1,\theta_2]\!]=[\![V.\theta_1,\theta_2]\!]+[\![\theta_1,V.\theta_2]\!]
\]
Taking into account
\[ [\![\theta^{can},\theta^{can}]\!]=V.[\![\theta^{can},\theta^{can}]\!]=\ldots=
V^n.[\theta^{can},\theta^{can}]\!]=0\]
we get:
\begin{eqnarray}
&& [\![\theta^{[2]},\theta^{[2]}]\!]=0 \\
&& [\![\theta^{[2]},\theta^{[0]}]\!]=0 \\
&& [\![\theta^{[0]},\theta^{[0]}]\!]+2[\![\theta^{[2]},\theta^{[-2]}]\!]=0 \\
&& [\![\theta^{[-2]},\theta^{[0]}]\!]=0 \\
&& [\![\theta^{[-2]},\theta^{[-2]}]\!]=0
\end{eqnarray}
We will verify explicitly that $\theta^{[0]}$ satisfies the Jacobi identity
$[\![\theta^{[0]},\theta^{[0]}]\!]=0$. This means that $\theta^{[2]}$, $\theta^{[0]}$
and $\theta^{[-2]}$ are three mutually compatible Poisson brackets.
This way of obtaining compatible Poisson brackets is similar to 
\cite{MR2079462}.

\section{Classical string on ${\bf R}\times S^N$, the nonlinear
$\sigma$-model and the generalized sine-Gordon system}
\label{sec:NLSM}
\subsection{Classical string and nonlinear $\sigma$-model}
Consider the classical string propagating on ${\bf R}\times S^N$,
where the first factor $\bf R$ is the time, and $S^N$ is  the space. 
 Let ${\bf x}=(x^1,\ldots, x^{N+1})$
be the unit vector parametrizing $S^N$, and let $T$ denote the time (the
coordinate parametrizing ${\bf R}$ in ${\bf R}\times S^N$).
We introduce on the string worldsheet the special set of coordinates
$(\tau^+,\tau^-)$ which are known as ``conformal coordinates''.
They are characterized by the Virasoro constraint:
\begin{eqnarray}
&&	\left({\partial t\over \partial \tau^+}\right)^2-
	\left({\partial {\bf x}\over \partial\tau^+}\right)^2=0\\[5pt]
&&	\left({\partial t\over \partial \tau^-}\right)^2-
	\left({\partial {\bf x}\over \partial\tau^-}\right)^2=0
\end{eqnarray}
Moreover we can choose $(\tau^+,\tau^-)$ so that:
\begin{equation}
t=\tau^++\tau^-
\end{equation}
Then we have:
\begin{equation}\label{NLSMwithVirasoroConstraints}
	\left({\partial {\bf x}\over \partial\tau^+}\right)^2=
	\left({\partial {\bf x}\over \partial\tau^-}\right)^2=1
\end{equation}
With these coordinates the equations of motion (the condition that the
worldsheet is an extremal surface) become the wave equations:
\begin{equation}\label{NLSMequation}
	D_{\bar{0}+} \partial_-{\bf x}=0
\end{equation}
Here $D_{\bar{0}}$ is the standard (Levi-Civita)
connection in the tangent space to the sphere:
\begin{equation}
	D_{\bar{0}+}V^i=\partial_+V^i+\Gamma^i_{jk}\partial_+ x^j V^k
\end{equation}
The index $\bar{0}$ is used for the consistency with the later notations. 
Equation (\ref{NLSMequation}) is called the ``nonlinear sigma-model'' (NLSM).
 The action of the nonlinear sigma-model follows from the Polyakov action
 of the classical string:
\begin{equation}\label{PolyakovAction}
\int d\tau^+ d\tau^- (\partial_+{\bf x},\partial_-{\bf x})
\end{equation}
The corresponding symplectic structure is
\begin{eqnarray}
	\Omega & =&\spacemaker{-}
	\int d\tau^+ (\delta {\bf x}, D_{\bar{0}+} \delta{\bf x})- \nonumber \\[5pt]
	&&	-\int d\tau^- (\delta {\bf x}, D_{\bar{0}-} \delta{\bf x})
\end{eqnarray}

\noindent
It is convenient to describe the string worldsheet 
in terms of the  group valued function $g(\tau^+,\tau^-)\in SO(N+1)$
modulo some gauge symmetry.

\begin{centering}
\begin{minipage}[t]{0.64\linewidth}
	\vspace{1pt}
\noindent
Let us pick some (constant) unit vector ${\bf x}_0\in S^N$.
Let $g^{-1}$ be the orthogonal matrix which rotates  
${\bf x}_0\in S^N$ to ${\bf x}(\tau^+,\tau^-)$. We have 
${\bf x}=g^{-1}{\bf x}_0$.
 Notice that $g$ is defined up to $g\simeq g_0 g$
where $g_0\in SO(N)$. This corresponds to the  gauge transformation
$g\simeq g_0g$.
The constant right shift 
$ g\mapsto gC,\;\;\; C\in SO(N+1)$, $C=\mbox{const} $
corresponds to the global rotations of $S^N$.
\end{minipage}
\hfill
\begin{minipage}[t]{0.32\linewidth}
\hspace{20pt}
\raisebox{-1.3in}{
\includegraphics[width=1.2in]{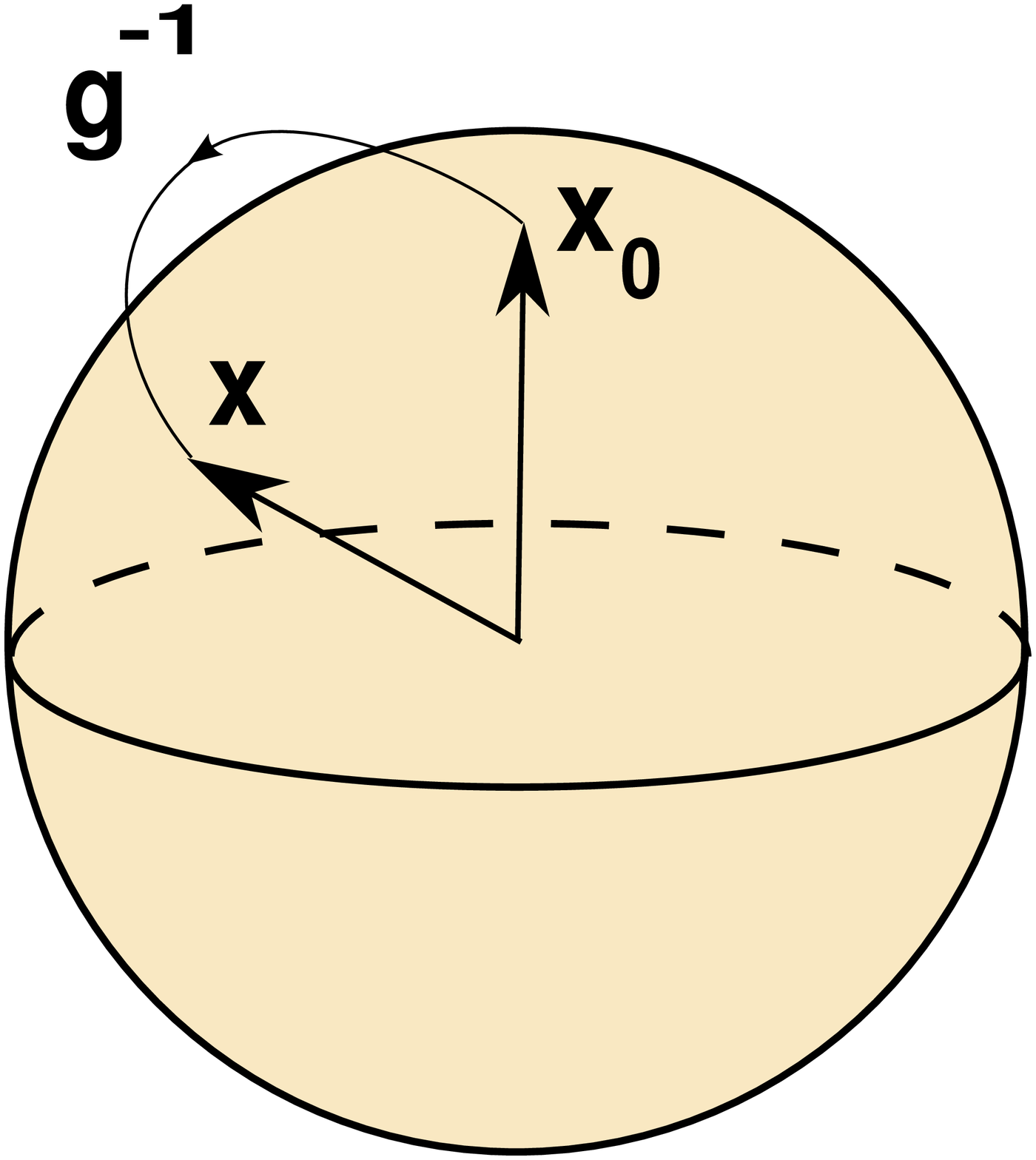}}
\end{minipage}\hfill
\end{centering}

\vspace{9pt}
\noindent
Therefore we can describe the string worldsheet in terms
of  $g(\tau^+,\tau^-)$ modulo the gauge invariance
\[
g(\tau^+,\tau^-)\equiv g_0(\tau^+,\tau^-)g(\tau^+,\tau^-),\;\;\;
g\in SO(N+1),\; g_0\in SO(N)
\]
\subsection{Classical string and generalized sine-Gordon}
We now want to rewrite the action (\ref{PolyakovAction}) in terms of $g$.
 Consider the Lie algebras
${\bf g}=so(N+1)$ and ${\bf g}_{\bar{0}}=so(N)$.
Introduce the ${\bf Z}_2$-grading\footnote{We use the notation ${\bf g}_{\bar{2}}$
rather than ${\bf g}_{\bar{1}}$ for the orthogonal complement
of ${\bf g}_{\bar{0}}\subset {\bf g}$, because we
want to ``leave some space'' for the odd generators
which will appear in the next section.
For the superstring we will have ${\bf g}={\bf g}_{\bar{0}}\oplus
{\bf g}_{\bar{1}}\oplus{\bf g}_{\bar{2}}\oplus{\bf g}_{\bar{3}}$
where ${\bf g}_{\bar{0}}\oplus {\bf g}_{\bar{2}}$ will be 
the even part of the superalgebra
and ${\bf g}_{\bar{1}}\oplus{\bf g}_{\bar{3}}$ the odd part.}
${\bf g}={\bf g}_{\bar{0}}\oplus 
{\bf g}_{\bar{2}}$:
{\footnotesize
\[
{\bf g}_{\bar{0}}: \; \left[
\begin{array}{cccccc}
0	&0	&0	&0	&0	&0	\\
0	&*	&*	&*	&*	&*	\\
0	&*	&*	&*	&*	&*	\\
0	&*	&*	&*	&*	&*	\\
0	&*	&*	&*	&*	&*	\\
0	&*	&*	&*	&*	&*	
\end{array}\right]
\; \; \; \;\;\;\;\;\;
{\bf g}_{\bar{2}}:\; \left[
\begin{array}{cccccc}
0	&*	&*	&*	&*	&*	\\{}
*	&0	&0	&0	&0	&0	\\{}
*	&0	&0	&0	&0	&0	\\{}
*	&0	&0	&0	&0	&0	\\{}
*	&0	&0	&0	&0	&0	\\{}
*	&0	&0	&0	&0	&0	
\end{array}\right]
\]
}
For an element $\xi\in {\bf g}$ we introduce the notation
\[
\xi=\xi_{\bar{0}}+\xi_{\bar{2}},\;\;\;
\xi_{\bar{0}}\in {\bf g}_0,\;\;\; \xi_{\bar{2}}\in {\bf g}_{\bar{2}}
\]
Also introduce the ``currents'' $J_{\pm}$:
\begin{equation}\label{DefineCurrents}
J_{\pm}=-\partial_{\pm}g g^{-1}
\end{equation}
The one-form current is defined as
\[
J=-dgg^{-1}=J_+d\tau^++J_-d\tau^-
\]
With these notations the action is 
\[S=-\int d\tau^+ d\tau^- \mbox{tr}\;J_{\bar{2}+}J_{\bar{2}-}\]
The Virasoro constraints are
\[
-\mbox{tr}(J_{\bar{2}+})^2=-\mbox{tr}(J_{\bar{2}-})^2=1
\]
We also have to remember that because of the definition 
(\ref{DefineCurrents}) the currents satisfy the Maurer-Cartan equation:
\begin{equation}
	dJ+J\wedge J=0
\end{equation}
The symplectic structure is
\begin{equation}
\Omega = \oint \left[d\tau^+\mbox{tr}\;(\delta g g^{-1})_{\bar{2}} 
	D_{ \bar{0} +}(\delta g g^{-1})_{\bar{2}}\;-(+\leftrightarrow -)\right]
\label{SymplecticNLSM}
\end{equation}
where $D_{\bar{0} +}=\partial_++\mbox{ad}_{J_{\bar{0}+}}$.
The equations of motion are:
\begin{eqnarray}
&&	\partial_+ J_{\bar{2}-}+ [J_{\bar{0}+},J_{\bar{2}-}]=0 \label{FirstGSG}\\
&&	\partial_- J_{\bar{2}+} + [J_{\bar{0}-},J_{\bar{2}+}] = 0 \label{SecondGSG}\\
&&	\partial_+ J_{\bar{0}-}-\partial_- J_{\bar{0}+}+
[J_{\bar{0}+},J_{\bar{0}-}]+[J_{\bar{2}+},J_{\bar{2}-}]=0 \label{ThirdGSG}
\end{eqnarray}
Notice that the equations of motion are written only in terms of $J$.
The group-valued field $g$ is related to $J$ by $J=-dgg^{-1}$, but the equations
of motion do not explicitly involve $g$.
The "generalized sine-Gordon" is obtained by forgetting about $g$.
Consider the space of solutions of the differential equations
(\ref{FirstGSG}), (\ref{SecondGSG}) and (\ref{ThirdGSG})
with the gauge symmetry 
\begin{equation}\label{GaugeSymmetryOfGSG}
\delta J= d\xi_0+[J,\xi_0],\;\;\; \xi_0\in {\bf g}_{\bar{0}}
\end{equation}
and the constraint
\begin{equation}\label{VirasoroAgain}
\mbox{tr}(J_{\bar{2}+})^2=\mbox{tr}(J_{\bar{2}-})^2=-1
\end{equation}
{\bf Definition.} The system of equations 
(\ref{FirstGSG}),(\ref{SecondGSG}) and (\ref{ThirdGSG}) with the gauge symmetry 
(\ref{GaugeSymmetryOfGSG}) and the constraint (\ref{VirasoroAgain}) is called  the
{\em generalized sine-Gordon (GSG)}. 

We get the usual sine-Gordon model when $N=2$, for the two-dimensional sphere.
In this case we can choose the gauge so that $J$ has the following form:
{\footnotesize
\begin{equation}
	J_+ =		\left[\begin{array}{ccc}
	0 		& \cos\varphi 		& \sin \varphi 		\\
	-\cos\varphi	& 0			& -\partial_+\varphi	\\
	-\sin\varphi	& \partial_+\varphi	& 0
			\end{array}\right]
\;\;\; , \;\;\;\;\;\;\;\;\;
	J_- =		\left[\begin{array}{ccc}
	0 		& \cos\varphi 		& -\sin \varphi 		\\
	-\cos\varphi	& 0			& \partial_-\varphi	\\
	\sin\varphi	& -\partial_-\varphi	& 0
			\end{array}\right]
\end{equation}
}
This solves Eqs. (\ref{FirstGSG}) and (\ref{SecondGSG}). Eq. (\ref{ThirdGSG})
leads to the usual sine-Gordon equation for $\varphi$:
\begin{equation}
	\partial_+\partial_-\varphi=-{1\over 2}\sin 2\varphi
\end{equation}
In some sense, the generalized sine-Gordon is equivalent to the nonlinear $\sigma$-model.
One only has to add $g$ satisfying $(d+J)g=0$.
But this $g$ is almost defined in terms of $J$, the only ambiguity comes from
the integration constants. (Which correspond to $g\mapsto gC$,
$C=\mbox{const}$, {\it i.e.} the global rotations of $S^{N}$.)

The equations of motion of the generalized sine-Gordon can be written as a zero curvature
equation if we introduce the {\em spectral parameter} $z$. Consider the operators:
\begin{eqnarray}
&&	{\cal L}_+[z]=\partial_++J_{\bar{0}+}+{1\over z^2} J_{\bar{2}+} \\
&&	{\cal L}_-[z]=\partial_-+J_{\bar{0}-}+z^2 J_{\bar{2}-}
\end{eqnarray}
The GSG equations of motion are equivalent to:
\begin{equation}
[{\cal L}_+,{\cal L}_-]=0
\end{equation}
In terms of $J$ the symplectic structure
Eq. (\ref{SymplecticNLSM})  is nonlocal:
\[
	\Omega  = 
	\oint d\tau^+\mbox{tr}\left((D_+^{-1}\delta J_+)_{\bar{2}}D_{\bar{0}+}
	(D_+^{-1}\delta J_+)_{\bar{2}}\right)-(+\leftrightarrow -) 
\]
But if we add $g$ satisfying $(d+J)g=0$ we get the local formula because
\begin{equation}\label{SomethingLikeG}
D_+^{-1} \delta J_+=-\delta g g^{-1}
\end{equation}
\[
	\Omega  = 
	\int d\tau^+\mbox{tr}\left((\delta g g^{-1})_{\bar{2}}D_{\bar{0}+}
	(\delta g g^{-1})_{\bar{2}}\right)-(+\leftrightarrow -) 
\]
The generalized sine-Gordon itself does not have a local symplectic structure
except for the special cases $N=2$ and $N=3$. 
To get the local symplectic structure, on has to slightly extend
the model by adding 
finitely many degrees of freedom. 
For example\footnote{
Another possibility would be to add the $\bf g$-valued
field $\Psi$ satisfying
$
D\Psi=*J_{\bar{2}}
$.
We would then get the symplectic structure:
$
\Omega=\oint \delta \Psi \delta J
$.
This corresponds to the action 
$
S=\int \mbox{tr}\left(\Psi(dJ+J^2)+ J_{\bar{2}}\wedge *J_{\bar{2}}\right)
$
The equation of motion for $\Psi$ implies the existence of $g$ such that
$J=-dgg^{-1}$ and the action on-shell is equal to 
the standard action 
$\int d\tau^+d\tau^- \mbox{tr}\;\left((\partial_+ g g^{-1})_{\bar{2}}
(\partial_- g g^{-1})_{\bar{2}}\right)$ and therefore gives essentially the 
same symplectic structure, modulo subtleties with boundary
conditions. 
Notice that the expression for $g$ in terms of $J$ is nonlocal,
so it is strictly
speaking a different model.
It could be thought of as a "T-dual"  
of the classical string \cite{MR1335189,MR2085789}.
}
adding $g$ 
satisfying $(d+J)g=0$ we get the canonical symplectic structure of the
nonlinear $\sigma$-model. In Section \ref{sec:Geometrical}
we will see that we can add some group-valued fields $g^L$ and $g^R$
satisfying (\ref{EquationForGLandGR}) and get the non-standard
symplectic structure (\ref{Omega}).

\subsection{Relativistic symmetry}
There is an obvious symmetry under the constant shifts of $\tau^+$ and $\tau^-$.
But besides shifts, the GSG equations are also symmetric under 
{\em boosts}:
\begin{eqnarray}
&&	J_{\bar{0}\pm}(\tau^+,\tau^-)\mapsto 
\lambda^{\pm 1} J_{\bar{0}\pm}(\lambda\tau^+,\lambda^{-1}\tau^-) \nonumber\\
&&	J_{\bar{2}\pm}(\tau^+,\tau^-)\mapsto 
J_{\bar{2}\pm}(\lambda\tau^+,\lambda^{-1}\tau^-) \nonumber
\end{eqnarray}
This  can be thought of as the rescaling of $(\tau_+,\tau_-)\mapsto (\lambda\tau^+,
\lambda^{-1}\tau^-)$ combined with the rescaling of the spectral parameter
$z\mapsto \lambda^{-1/2}z$. 

We will use this relativistic symmetry to introduce the {\em bihamiltonian structure}
of the GSG. It turns out that the canonical Poisson structure of the NLSM leads
to the Poisson structure $\theta^{str}$ of the GSG which is not invariant
under the relativistic symmetry. More precisely we will have:
\begin{equation}
\theta_{str}=\theta^{[2]}+\theta^{[0]}+\theta^{[-2]}
\end{equation}
where $\theta^{[0]}$ is invariant under the boosts, and $\theta^{[2]}$ and 
$\theta^{[-2]}$ transform as $\lambda^2$ and $\lambda^{-2}$ respectively.
We will explicitly verify that $\theta^{[0]}$ satisfies the Jacobi
identities. Then the arguments of Section \ref{sec:Compatibility}
imply that  $\theta^{[2]}$ and $\theta^{[-2]}$ also satisfy
the Jacobi identities, and moreover all three brackets
$\theta^{[0]}$, $\theta^{[2]}$ and $\theta^{[-2]}$ are mutually compatible. 
We will also find that $\theta^{[-2]}$ is related to
$\theta^{[0]}$ and $\theta^{[2]}$ by Eq. (\ref{RelationBetweenBrackets}).

\section{Superstring in $AdS_5\times S^5$}
\label{sec:Superstring}
In this section we will use a variant of the Metsaev-Tseytlin description
of the superstring in $AdS_5\times S^5$ \cite{MetsaevTseytlin}. 

\subsection{The superalgebra ${\bf g}=psu(2,2|4)$}
One of the most important properties of this 
superalgebra is the existence of a  ${\bf Z}_4$ grading
\cite{MetsaevTseytlin,RoibanSiegel,Bena:2003wd,Beisert:2005bm}.
There are many ways to explain this grading. 
For example, we can use the correspondence between the
bosonic generators of the superalgebra and the Killing vector
fields on $AdS_5\times S^5$. The fermionic generators correspond
to Killing spinors. 
We can embed $AdS_5\times S^5$ into the flat space ${\bf R}^{2+10}$
as the direct product of the hyperboloid and the sphere. 
As explained in \cite{BarKillingSpinors,Mikhailov:2000ya}
the Killing spinors in $AdS_5\times S^5$ correspond to the constant
spinors in ${\bf R}^{2+10}$ satisfying some chirality condition.  
The spinor bundle of $AdS_5\times S^5$ is naturally identified
with the subbundle of the spinor bundle of ${\bf R}^{2+10}$ which
is the image of the projector 
\[
{1\over 2}\left(1+\Gamma(e^{\perp}_A)\Gamma(e^{\perp}_S)\right)
\]
Here $e^{\perp}_S$ is the vector field normal to the surface
of $S^5$ in ${\bf R}^6$, and $e^{\perp}_A$ is normal to the
surface of the hyperboloid in ${\bf R}^{2+4}$.
For any vector $v$ we denote 
$\Gamma(v)$ the corresponding $\Gamma$-matrix
$\Gamma_{\mu}v^{\mu}$. We assume that $\Gamma_{-1}^2=\Gamma_0^2=1$
and $\Gamma_1^2=\ldots=\Gamma_{10}^2=-1$.
The covariantly constant spinors corresponds to the sections 
of the form
\begin{equation}\label{psi}
\psi={1\over 2}\left(1+\Gamma(e^{\perp}_A)\Gamma(e^{\perp}_S)\right)
\Psi_{++}
\end{equation}
where $\Psi_{++}$ is a constant spinor with the chirality conditions:
\begin{eqnarray}
	\Gamma_{-1}\Gamma_0\Gamma_1\Gamma_2\Gamma_3\Gamma_4 \Psi_{++}=i\Psi_{++}
	\label{ChiralityAdS}
\\
	\Gamma_5\Gamma_6\Gamma_7\Gamma_8\Gamma_9\Gamma_{10}\Psi_{++}=i\Psi_{++}
	\label{ChiralitySphere}
\end{eqnarray}
In this situation we will write:
\begin{equation}
	\Psi_{++}={\cal S}(\psi)
\end{equation}
We defined $\cal S$ as the map (\ref{psi}) 
from the spinors in $AdS_5\times S^5$
to the spinors in ${\bf R}^{2+10}$ with the chirality conditions
(\ref{ChiralityAdS}),(\ref{ChiralitySphere}).
The ${\bf Z}_4$ grading depends on the choice of a point 
$x_0\in AdS_5\times S^5$. Let $e_{0A}^{\perp}$ and $e_{0S}^{\perp}$
be the corresponding unit normals at the point $x_0$. 
The ${\bf Z}_4$ grading is defined by the operator $\Lambda$:
\begin{equation}
	\Lambda\Psi_{++}=\Gamma(e_{0A}^{\perp})\Gamma(e_{0S}^{\perp})\Psi_{++}^*
\end{equation}
Notice that $\Lambda^2={\bf 1}$.
We will say that $\psi$ given by (\ref{psi}) belongs to ${\bf g}_{\bar{1}}$
if $\Lambda\Psi_{++}=\Psi_{++}$ and to ${\bf g}_{\bar{3}}$ if
$\Lambda\Psi_{++}=-\Psi_{++}$.
Here $*$ means complex conjugation, and we use such a representation
of the $\Gamma$-matrices $\Gamma_{-1},\ldots,\Gamma_{10}$ that
all their components are real numbers. 
As in the bosonic case, ${\bf g}_{\bar{0}}$ is the stabilizer of
$x_0$ and ${\bf g}_{\bar{2}}$ is the bosonic part of the orthogonal
complement to this stabilizer.
\subsection{Classical action and the canonical symplectic structure}
For the superstring the current $J$ belongs to the superalgebra 
${\bf g}=psu(2,2|4)$ and can be decomposed according to its ${\bf Z}_4$
grading:
\begin{equation}
	J=J_{\bar{0}}+J_{\bar{1}}+J_{\bar{2}}+J_{\bar{3}}
\end{equation}
(For the purely bosonic string we had only the even components
$J_{\bar{0}}$ and $J_{\bar{2}}$.)

The classical action is
\begin{equation}
	S={1\over 2}\int\int \mbox{str}\left[
	J_{\bar{2}}\;*J_{\bar{2}}+J_{\bar{1}}J_{\bar{3}}\right]
\end{equation}
The Maurer-Cartan equations are:
\begin{eqnarray*}
&&	dJ_2+J_0J_2+J_2J_0+J_1^2+J_3^2=0 \\
&&	dJ_1+J_0J_1+J_1J_0+J_2J_3+J_3J_2=0 \\
&&	dJ_3+J_0J_3+J_3J_0+J_1J_2+J_2J_1=0
\end{eqnarray*}
The equations of motion are:
\begin{eqnarray*}
&&	d*J_2+J_0*J_2+*J_2J_0-J_1^2+J_3^2=0	\\
&&	[J_{3-},J_{2+}]=0	\\
&&	[J_{1+},J_{2-}]=0
\end{eqnarray*}
The symplectic potential (\ref{DefSymplecticPotential})
follows from the on-shell variation of the action:
\begin{equation}
	\alpha={1\over 2}\int \mbox{str}\left(2(\delta g g^{-1})_{\bar{2}}
	*J_{\bar{2}}+(\delta g g^{-1})_{\bar{1}} J_{\bar{3}}
	-(\delta g g^{-1})_{\bar{3}}J_{\bar{1}}\right)
\end{equation}
The symplectic form is $\Omega=\delta\alpha$:
\begin{eqnarray}
\Omega & =\int \mbox{str} & \left\{ -(\delta g g^{-1})_{\bar{2}} \wedge
*{D}_{J_{\bar{0}}}(\delta g g^{-1})_{\bar{2}}
+\right.\nonumber\\
&&	+(J_{\bar{2}}-*J_{\bar{2}})
	(\delta g g^{-1})_{\bar{1}}\wedge (\delta g g^{-1})_{\bar{1}}-
	\\
&&	-(J_{\bar{2}}+*J_{\bar{2}})
	(\delta g g^{-1})_{\bar{3}}\wedge (\delta g g^{-1})_{\bar{3}}+
	\nonumber \\
	&&	+(J_{\bar{1}}+*J_{\bar{1}}) 
	( (\delta g g^{-1})_{\bar{1}}\wedge (\delta g g^{-1})_{\bar{2}}+
	  (\delta g g^{-1})_{\bar{2}}\wedge (\delta g g^{-1})_{\bar{1}} )-
	\nonumber \\
&&	\left.
	-(J_{\bar{3}}-*J_{\bar{3}})
( (\delta g g^{-1})_{\bar{3}}\wedge (\delta g g^{-1})_{\bar{2}}
 +(\delta g g^{-1})_{\bar{2}}\wedge (\delta g g^{-1})_{\bar{3}})\right\}
	\nonumber
\end{eqnarray}
This form is  strictly speaking not symplectic, because the variations
\begin{equation}\label{Kernel}
\delta g g^{-1}=f^+(\tau^+,\tau^-)J_{\bar{1}+}+
f^-(\tau^+,\tau^-)J_{\bar{3}-}
\end{equation}
are in the kernel of
$\Omega$ for an arbitrary $f^{\pm}(\tau^+,\tau^-)$. The symplectic
form by definition should be nondegenerate; we should have called $\Omega$
``presymplectic''.
The variation (\ref{Kernel}) should therefore be considered a gauge
transformation.
It
preserves the equations of motion.

\vspace{10pt}
{\small
\noindent
The situation in flat space is similar, but technically simpler.
There are two fermions $\theta^1$ and $\theta^2$.
Let us restrict ourselves with the quadratic order, in the fermions.
The  currents $J_{\bar{1}}$ and $J_{\bar{3}}$ correspond to 
$d\theta^1$ and $d\theta^2$. The equations of motion 
$[J_{\bar{2}+},J_{\bar{3}-}]=0$ and $[J_{\bar{2}-},J_{\bar{1}+}]=0$
correspond to 
\begin{equation}\label{FlatEqM}
	\widehat{\partial_+ x}\partial_-\theta^2=
	\widehat{\partial_- x}\partial_+\theta^1=0
\end{equation}
The symplectic form (assuming $\widehat{\partial_+ x}$
and $\widehat{\partial_-x}$ constant) is 
\begin{equation}\label{FlatOmega}
\Omega=	\int d\tau^- \delta \overline{\theta^1}\;
\widehat{\partial_-x}\; \delta \theta^1 + 
	\int d\tau^+ \delta \overline{\theta^2}\;
\widehat{\partial_+x}\; \delta \theta^2
\end{equation}
The kernel is $\delta_f \theta^1 = f^+\partial_+\theta^1$ and
$\delta_f \theta^2 = f^- \partial_- \theta^2$ (to verify
that this is in the kernel of (\ref{FlatOmega}), we 
have to use (\ref{FlatEqM})). But this is actually a kappa-symmetry.
There are such $\kappa^1$ and $\kappa^2$ that 
$\delta_f \theta^1= \widehat{\partial_- x} \kappa^1$ and
$\delta_f \theta^2= \widehat{\partial_+ x} \kappa^2$.
On the other hand, there is a gauge with 
$\partial_+\theta^1=\partial_-\theta^2=0$. (We used this gauge in
\cite{Mikhailov:2004xw}.)
}

\vspace{10pt}
\noindent
We will put 
\begin{equation}\label{SomeFermionicCurrentsZero}
	J_{\bar{1}+}=J_{\bar{3}-}=0
\end{equation}
This gives a gauge-fixed version of the Metsaev-Tseytlin formalism.
The equations of motion become:
\begin{eqnarray}
	&&	D_{\bar{0}+}J_{\bar{2}-}=D_{\bar{0}-}J_{\bar{2}+}=0
	\label{MTFirstEqM}\\
	&&	(D_+J_-)_{\bar{1}}=(D_-J_+)_{\bar{3}}=0
	\label{MTSecondEqM}
\end{eqnarray}
This can be understood as the consistency condition for the zero curvature
equation:
\begin{equation}
	\left[
	D_{\bar{0}+}+{1\over z}J_{\bar{3}+}+{1\over z^2}J_{\bar{2}+}\; , \;
	D_{0-}+zJ_{\bar{1}-}+z^2J_{\bar{2}-}
	\right] = 0
\end{equation}
Eq. (\ref{SomeFermionicCurrentsZero}) implies the
following constraint on $\xi=\delta g g^{-1}$:
\begin{eqnarray}
	&&	D_{\bar{0}+}\xi_{\bar{1}}+
	[J_{\bar{2}+},\xi_{\bar{3}}]+[J_{\bar{3}+},\xi_{\bar{2}}]=0
	\label{XiOne}\\
	&&	D_{\bar{0}-}\xi_{\bar{3}}+[J_{\bar{2}-},\xi_{\bar{1}}]+
	[J_{\bar{1}-},\xi_{\bar{2}}]=0
\end{eqnarray}
In the classical string worldsheet theory $J$ should satisfy the
Virasoro constraint:
\begin{equation}\label{VirasoroConstraintOnJTwo}
\mbox{str}\; J_{\bar{2}+}^2 =0
\end{equation}
In what follows we will assume that the Virasoro constraints are
satisfied.

\subsection{Gauge transformations and dressing transformations}
Let us consider the left invariant vector fields $L_{\xi}$ such
that $L_{\xi}.g=-\xi g$. 
The symplectic form is:
\begin{eqnarray}
\Omega(L_{\xi},L_{\eta}) &=&
\int d\tau^+ \mbox{str}\left(\eta_{\bar{2}} 
\stackrel{\leftrightarrow}{D}_{ \bar{0}+}\xi_{ \bar{2}} - 
\eta_{\bar{3}}\stackrel{\leftrightarrow}{ad}_{J_{\bar{2}+}}\xi_{\bar{3}}\right)
-\label{SymplecticForm}\\
&-&	\int d\tau^-
\mbox{str}\left(\eta_{\bar{2}} 
\stackrel{\leftrightarrow}{D}_{\bar{0}-}\xi_{\bar{2}} - 
\eta_{\bar{1} }\stackrel{\leftrightarrow}{ad}_{J_{\bar{2}-}}\xi_{\bar{1}}
\right)
\label{SymplecticFormSecondLine}
\end{eqnarray}
Strictly speaking, this is not yet a symplectic form, 
because it still has a kernel.
The kernel is generated by the vectors of the form
$(\xi_{\bar{0}},\xi_{\bar{1}},\xi_{\bar{2}},\xi_{\bar{3}})
=(0,0,0,\chi_{\bar{3}})$ where $\chi_{\bar{3}}$ is such that
$[J_{\bar{2}+},\chi_{\bar{3}}]=0$ and 
$D_{\bar{0}-}\chi_{\bar{3}}=0$, and the vectors of the form
$(\xi_{\bar{0}},\xi_{\bar{1}},\xi_{\bar{2}},\xi_{\bar{3}})=
(0,\chi_{\bar{1}},0,0)$ where $\chi_{\bar{1}}$ is such that
$[J_{\bar{2}-},\chi_{\bar{1}}]=0$ and $D_{\bar{0}+}\chi_{\bar{1}}=0$.
These are the residual 
gauge transformations.
These gauge transformations can be considered as particular dressing
transformations with the parameter 
$\chi(z)=z^{-1}\chi_{\bar{3}}+z\chi_{\bar{1}}$:
\begin{eqnarray}
&&	\delta J(z) = [d+J(z),\chi(z)] \label{GaugeDressing} \\[5pt]
&&	\chi(z)=z^{-1}\chi_{\bar{3}}+z\chi_{\bar{1}} \\[5pt]
&&	[J_{\bar{2}-},\chi_{\bar{1}}]=0,\;\;\; D_{\bar{0}+}\chi_{\bar{1}}=0\\
&&	[J_{\bar{2}+},\chi_{\bar{3}}]=0,\;\;\; D_{\bar{0}-}\chi_{\bar{3}}=0
\end{eqnarray}
In the lightcone formalism, the independent variables are $J_+$.
We will fix these residual gauge transformations by requiring 
that there exists $K_{\bar{1}}$ such that
\begin{equation}\label{SpecialGaugeChoiceFermions}
	J_{\bar{3}+}=[J_{\bar{2}+}, K_{\bar{1}}]
\end{equation}
In terms of the 12-dimensional spinors:
\begin{equation}
	{\cal S}(\hat{J}_{\bar{3}+})=
	\left(	\widehat{\partial_+x}_S\Gamma(e^{\perp}_S)+
	\widehat{\partial_+x}_A\Gamma(e^{\perp}_A)	
	\right)	{\cal S}(K_{\bar{1}})
\end{equation}
We have to explain why (\ref{SpecialGaugeChoiceFermions})
is a reasonable gauge choice.
Consider the projection operator
\begin{equation}
	{\cal P}={1\over 2}\left(1-
	\widehat{\partial_+x}_S\Gamma(e^{\perp}_S)
	\widehat{\partial_+x}_A\Gamma(e^{\perp}_A)\right)
\end{equation}
Notice that ${\cal P}^2={\bf 1}$.
We have:
\begin{equation}
	\begin{array}{l}
\mbox{Ker}(\mbox{ad}_{J_{\bar{2}+}}:{\bf g}_{\bar{3}}\rightarrow
{\bf g}_{\bar{1}})= \mbox{Ker }{\cal P} \\
\mbox{Im}( \mbox{ad}_{J_{\bar{2}+}}:{\bf g}_{\bar{1}}\rightarrow
{\bf g}_{\bar{3}})	= \mbox{Im }{\cal P}
\end{array}
\end{equation}
Now suppose that
\begin{equation}
	J_{\bar{3}+}=[J_{\bar{2}+},K_{\bar{1}}]+\Delta J_{\bar{3}+}
\end{equation}
where $\Delta J_{\bar{3}_+}$ is small and belongs to 
$\mbox{Ker}(\mbox{ad}_{J_{\bar{2}+}}:{\bf g}_{\bar{3}}\rightarrow
{\bf g}_{\bar{1}})$.
We want to prove that there is a small 
$\chi_{\bar{3}}\in \mbox{Ker}
(\mbox{ad}_{J_{\bar{2}+}}:{\bf g}_{\bar{3}}\rightarrow
{\bf g}_{\bar{1}})$ and $\Delta K_{\bar{1}}$ such that
\begin{equation}
	D_{\bar{0}+}\chi_{\bar{3}}-\mbox{ad}_{J_{\bar{2}+}}\Delta K_{\bar{1}}
	=\Delta J_{\bar{3}+}
\end{equation}
This means that we are looking for $\chi_{\bar{3}}$ such that:
\begin{equation}
	(1-{\cal P})D_{\bar{0}+}\chi_{\bar{3}}=\Delta J_{\bar{3}+}
\end{equation}
Therefore we have to prove that the operator 
\[
{\cal A}=(1-{\cal P})D_{\bar{0}+}: \mbox{Ker}{\cal P}\rightarrow 
\mbox{Ker}{\cal P}
\]
is invertible. 
This is true when $J_{\bar{0}+}=0$ and $J_{\bar{2}+}=\mbox{const}$
because in this case
we have $\left.(1-{\cal P})D_{\bar{0}+}\right|_{\mbox{Ker}{\cal P}}
=\partial_+$.
This means that ${\cal A}$ will remain invertible at least for
small enough $J_{\bar{0}+}$ and slowly varying $J_{\bar{2}+}$.

\section{Bihamiltonian structure of the classical superstring}
\label{sec:Bihamiltonian}
\subsection{Hidden relativistic symmetry}
\label{sec:HiddenRelativisticSymmetry}
The relativistic symmetry acts in the following way:
\begin{eqnarray}
&&	J_{\bar{0}\pm}(\tau^+,\tau^-)\mapsto 
\lambda^{\pm 1} J_{\bar{0}\pm}(\lambda\tau^+,\lambda^{-1}\tau^-) \nonumber\\
&&	J_{\bar{1}-}(\tau^+,\tau^-)\mapsto 
\lambda^{- 1/2} J_{\bar{1}-}(\lambda\tau^+,\lambda^{-1}\tau^-)
\nonumber \\
&&	J_{\bar{3}+}(\tau^+,\tau^-)\mapsto 
\lambda^{ 1/2} J_{\bar{3}+}(\lambda\tau^+,\lambda^{-1}\tau^-)
\label{HiddenRelativisticSymmetry} \\
&&	J_{\bar{2}\pm}(\tau^+,\tau^-)\mapsto 
J_{\bar{2}\pm}(\lambda\tau^+,\lambda^{-1}\tau^-)
\nonumber
\end{eqnarray}
This is a symmetry of the equations of motion 
(\ref{MTFirstEqM}),(\ref{MTSecondEqM}).
This  can be thought of as the rescaling of 
$(\tau_+,\tau_-)\mapsto (\lambda\tau^+,
\lambda^{-1}\tau^-)$ combined with the rescaling of the spectral parameter
$z\mapsto \lambda^{-1/2}z$. 

\subsection{The canonical Poisson bracket}
In this section we will calculate the Poisson bracket corresponding
to the symplectic form 
(\ref{SymplecticForm}), (\ref{SymplecticFormSecondLine}).
We use the lightcone formalism, so only the first line
(\ref{SymplecticForm}) is important for us. The integral over $d\tau^+$
is from $-\infty$ to $+\infty$, at constant $\tau^-$.

Let us formally resolve the constraint (\ref{XiOne}):
\begin{equation}\label{SolvedForXiOne}
	\xi_{\bar{1}}=-
	D_{\bar{0}+}^{-1}([J_{\bar{2}+},\xi_{\bar{3}}]+[J_{\bar{3}+},\xi_{\bar{2}}])
\end{equation}
Consider the variation with $\xi_{\bar{0}}=0$ and $\xi_{\bar{1}}$ given by 
(\ref{SolvedForXiOne}):
\begin{eqnarray}
&&\delta J_{\bar{3}+}=
D_{\bar{0}+}\xi_{\bar{3}}-[J_{\bar{2}+}, D_{\bar{0}+}^{-1}([J_{\bar{2}+},\xi_{\bar{3}}]
+[J_{\bar{3}+},\xi_{\bar{2}}])] \nonumber\\
&&	\delta J_{\bar{2}+}=
D_{\bar{0}+}\xi_{\bar{2}}+[J_{\bar{3}+},\xi_{\bar{3}}]
\label{DeltaJvsXi} \\
&&\delta J_{\bar{0}+}=
[J_{\bar{2}+},\xi_{\bar{2}}]-[J_{\bar{3}+},D_{\bar{0}+}^{-1}([J_{\bar{2}+},\xi_{\bar{3}}]
+[J_{3+},\xi_{\bar{2}}])]\nonumber
\end{eqnarray}
Consider a functional $F$, which is gauge invariant under $\delta J= [d+J,\xi_{\bar{0}}]$
and under (\ref{GaugeDressing}). Being invariant
under (\ref{GaugeDressing}) implies that:
\begin{eqnarray}
	\mbox{ad}_{J_{\bar{3}+}}{\delta F\over\delta J_{\bar{2}+}}
	+D_{\bar{0}+}{\delta F\over\delta J_{\bar{3}+}} 
	\in \mbox{Im}(\mbox{ad}_{J_{\bar{2}}}) \label{ImageOfAdjoint} \\
\mbox{ad}_{J_{\bar{3}+}}{\delta F\over\delta J_{\bar{0}+}}
+\mbox{ad}_{J_{\bar{2}+}}{\delta F\over\delta J_{\bar{3}+}} 
\in \mbox{Im}(D_{\bar{0}+})
\end{eqnarray}
Let us find $\eta$ such that the $L_{\eta}$ is  the Hamiltonian vector field
generated by $F$: 
\begin{equation}
\Omega(L_{\eta},L_{\xi})=L_{\xi}.F   \;\;\;\;\; \mbox{for any $\xi$ }
\end{equation}
Let us denote $a_{\bar{2}}=\mbox{ad}_{J_{\bar{2}+}}$ and
$a_{\bar{3}}=\mbox{ad}_{J_{\bar{3}+}}$.
 A straightforward calculation gives:
\begin{eqnarray*}
	\eta_{\bar{1}}&=&{\delta F\over\delta J_{\bar{3}+}}+
(( D_{\bar{0}+}^{-1}a_{\bar{3}} )^2 D_{\bar{0}+}^{-1}a_{\bar{2}} 
- (D_{\bar{0}+}^{-1}a_{\bar{2}})^2){\delta F\over\delta J_{\bar{3}+}}-\\
&&	-(D_{\bar{0}+}^{-1}a_{\bar{2}}D_{\bar{0}+}^{-1}a_{\bar{3}}+D_{\bar{0}+}^{-1}a_{\bar{3}}D_{\bar{0}+}^{-1}a_{\bar{2}}-
(D_{\bar{0}+}^{-1}a_{\bar{3}})^3)
{\delta F\over\delta J_{\bar{0}+}}
\\
	\eta_{\bar{2}}&=&{\delta F\over\delta J_{\bar{2}+}} -
D_{\bar{0}+}^{-1}a_{\bar{3}}D_{\bar{0}+}^{-1}a_{\bar{2}}{\delta F\over\delta J_{\bar{3}+}}
+
(D_{\bar{0}+}^{-1}a_{\bar{2}}-(D_{\bar{0}+}^{-1}a_{\bar{3}})^2){\delta F\over\delta J_{\bar{0}+}}
\\
	\eta_{\bar{3}}&=&(D_{\bar{0}+}^{-1}a_{\bar{2}}-a_{\bar{2}}^{-1}D_{\bar{0}+}){\delta F\over\delta J_{\bar{3}+}}-
a_{\bar{2}}^{-1}a_{\bar{3}}{\delta F\over\delta J_{\bar{2}+}}
+D_{\bar{0}+}^{-1}a_{\bar{3}}{\delta F\over\delta J_{\bar{0}+}}
\end{eqnarray*}
where we have denoted
\[
a_{\bar{2}}=\mbox{ad}_{J_{\bar{2}+}}\;,\;\;\;
a_{\bar{3}}=\mbox{ad}_{J_{\bar{3}+}}
\]
This gives  the Poisson bracket of $F$ with the currents:
\begin{eqnarray}
\{F, J_{\bar{2}+}\}&=&
	(D_{\bar{0}+}-a_{\bar{3}}a_{\bar{2}}^{-1}a_{\bar{3}}){\delta F\over \delta J_{\bar{2}+}}-
	a_{\bar{3}}a_{\bar{2}}^{-1}D_{\bar{0}+}{\delta F\over \delta J_{\bar{3}+}}+
	a_{\bar{2}}{\delta F\over\delta J_{\bar{0}+}} \nonumber\\[7pt]
\{F, J_{\bar{3}+} \}&=&
	(a_{\bar{3}}-(a_{\bar{2}}D_{\bar{0}+}^{-1})^2a_{\bar{3}}-
	a_{\bar{2}}D_{\bar{0}+}^{-1}a_{\bar{3}}D_{\bar{0}+}^{-1}a_{\bar{2}}+
	a_{\bar{2}}(D_{\bar{0}+}^{-1}a_{\bar{3}})^3)
	{\delta F\over\delta J_{\bar{0}+}}- \nonumber \\
&&	-D_{\bar{0}+}a_{\bar{2}}^{-1}a_{\bar{3}} {\delta F\over\delta J_{\bar{2}+}}+\nonumber\\
&&	+\left( -a_{\bar{2}}(a_{\bar{2}}^{-1}D_{\bar{0}+}-D_{\bar{0}+}^{-1}a_{\bar{2}})^2+
	a_{\bar{2}}(D_{\bar{0}+}^{-1}a_{\bar{3}})^2D_{\bar{0}+}^{-1}a_{\bar{2}}\right)
	{\delta F\over\delta J_{\bar{3}+}}
	\nonumber\\[7pt]
\{F, J_{\bar{0}+} \}&=&
	\left(-a_{\bar{3}}D_{\bar{0}+}^{-1}a_{\bar{2}}D_{\bar{0}+}^{-1}a_{\bar{3}}+
	a_{\bar{3}}(D_{\bar{0}+}^{-1}a_{\bar{3}})^3+ \right.\nonumber \\ 
&&	\left.+a_{\bar{2}}D_{\bar{0}+}^{-1}a_{\bar{2}}-
	a_{\bar{2}}(D_{\bar{0}+}^{-1}a_{\bar{3}})^2-
	(a_{\bar{3}}D_{\bar{0}+}^{-1})^2a_{\bar{2}}\right)
	{\delta F\over\delta J_{\bar{0}+}}+	\nonumber \\ 
&&	+a_{\bar{2}}{\delta F\over \delta J_{\bar{2}+}} +\nonumber \\
&&	+\left(a_{\bar{3}}-
	a_{\bar{2}} D_{\bar{0}+}^{-1} a_{\bar{3}} D_{\bar{0}+}^{-1} a_{\bar{2}} +
	(a_{\bar{3}}D_{\bar{0}+}^{-1})^3 a_{\bar{2}} - 
	a_{\bar{3}} (D_{\bar{0}+}^{-1}a_{\bar{2}})^2\right)
	{\delta F\over\delta J_{\bar{3}+}}
\nonumber
\end{eqnarray}
Therefore the Poisson bracket of the currents is:
\begin{eqnarray}
\{J_{\bar{2}+},J_{\bar{2}+}\}&=&
	-D_{\bar{0}+}+a_{\bar{3}}a_{\bar{2}}^{-1}a_{\bar{3}}\nonumber	\\[5pt]
\{J_{\bar{2}+},J_{\bar{3}+}\}&=&a_{\bar{3}}a_{\bar{2}}^{-1}D_{\bar{0}+}     \nonumber	\\[5pt]
\{J_{\bar{2}+},J_{\bar{0}+}\}&=&-a_{\bar{2}}\nonumber\\[5pt]
\{J_{\bar{3}+},J_{\bar{0}+}\}&=&
	-a_{\bar{3}}+a_{\bar{2}}D_{\bar{0}+}^{-1}a_{\bar{3}}D_{\bar{0}+}^{-1}a_{\bar{2}}-
a_{\bar{2}}D_{\bar{0}+}^{-1}a_{\bar{3}}D_{\bar{0}+}^{-1}a_{\bar{3}}D_{\bar{0}+}^{-1}a_{\bar{3}}+
	\nonumber\\
&&	+a_{\bar{2}}D_{\bar{0}+}^{-1}a_{\bar{2}}D_{\bar{0}+}^{-1}a_{\bar{3}}\nonumber\\[5pt]
\{J_{\bar{3}+},J_{\bar{3}+}\}&=&(D_{\bar{0}+}-a_{\bar{2}}D_{\bar{0}+}^{-1}a_{\bar{2}})
	a_{\bar{2}}^{-1}(D_{\bar{0}+}-a_{\bar{2}}D_{\bar{0}+}^{-1}a_{\bar{2}})-\nonumber \\
&& 	
-a_{\bar{2}}D_{\bar{0}+}^{-1}a_{\bar{3}}D_{\bar{0}+}^{-1}a_{\bar{3}}D_{\bar{0}+}^{-1}a_{\bar{2}}
	\nonumber\\[5pt]
\{J_{\bar{0}+},J_{\bar{0}+}\}&=&-a_{\bar{2}}D_{\bar{0}+}^{-1}a_{\bar{2}}+
	a_{\bar{3}}D_{\bar{0}+}^{-1}a_{\bar{3}}D_{\bar{0}+}^{-1}a_{\bar{2}}+
	a_{\bar{2}}D_{\bar{0}+}^{-1}a_{\bar{3}}D_{\bar{0}+}^{-1}a_{\bar{3}}+
	\nonumber\\
&& 	+a_{\bar{3}}D_{\bar{0}+}^{-1}a_{\bar{2}}D_{\bar{0}+}^{-1}a_{\bar{3}}-
a_{\bar{3}}D_{\bar{0}+}^{-1}a_{\bar{3}}D_{\bar{0}+}^{-1}a_{\bar{3}}D_{\bar{0}+}^{-1}a_{\bar{3}}
\nonumber
\end{eqnarray}
We should remember that these Poisson brackets are defined
only on those functions which are gauge invariant.
Notice that $a_{\bar{2}}^{-1}$ exists because of (\ref{ImageOfAdjoint}).

\subsection{Action of the boosts on the canonical Poisson bracket}
The canonical Poisson brackets are not invariant under the boosts. 
But it turns out tha the boosts act on the canonical bracket
in some relatively simple way.
In fact
 the canonical bracket is the sum of three
terms each having a definite scaling dimension: 
\begin{equation}
\theta_{str}=\theta^{[2]}+\theta^{[0]}+\theta^{[-2]}
\end{equation}
We have the following expressions for $\theta^{[2]}$ and
$\theta^{[0]}$. The lowest grade $\theta^{[2]}$ is:
\begin{eqnarray}
&&	\{J_{\bar{2}+},J_{\bar{2}+}\}^{[2]}=
-D_{\bar{0}+}+a_{\bar{3}}a_{\bar{2}}^{-1}a_{\bar{3}} \nonumber\\
&&	\{J_{\bar{2}+},J_{\bar{3}+}\}^{[2]}=
a_{\bar{3}}a_{\bar{2}}^{-1}D_{\bar{0+}} 
\label{ThetaTwo}	\\
&&	\{J_{\bar{3}+},J_{\bar{3}+}\}^{[2]}=D_{\bar{0}+}
a_{\bar{2}}^{-1}D_{\bar{0}+}
\nonumber
\end{eqnarray}
The next grade is $\theta^{[0]}$:
\begin{eqnarray}
&&	\{J_{\bar{2}+},J_{\bar{0}+}\}^{[0]}=-a_{\bar{2}} \nonumber\\
&&	\{J_{\bar{3}+},J_{\bar{0}+}\}^{[0]}=-a_{\bar{3}} \label{BoostInvariant}\\
&&	\{J_{\bar{3}+},J_{\bar{3}+}\}^{[0]}=-2a_{\bar{2}}\nonumber
\end{eqnarray}
And the highest grade is $\theta^{[-2]}$:
\begin{eqnarray}
	\{J_{\bar{3}+},J_{\bar{0}+}\}^{[-2]}&=&
a_{\bar{2}}D_{\bar{0}+}^{-1}a_{\bar{3}}D_{\bar{0}+}^{-1}a_{\bar{2}}-a_{\bar{2}}D_{\bar{0}+}^{-1}a_{\bar{3}}D_{\bar{0}+}^{-1}a_{\bar{3}}D_{\bar{0}+}^{-1}a_{\bar{3}}+
\nonumber\\
&&	+a_{\bar{2}}D_{\bar{0}+}^{-1}a_{\bar{2}}D_{\bar{0}+}^{-1}a_{\bar{3}} \nonumber \\[5pt]
	\{J_{\bar{3}+},J_{\bar{3}+}\}^{[-2]}&=&
a_{\bar{2}}D_{\bar{0}}^{-1}a_{\bar{2}}D_{\bar{0}}^{-1}a_{\bar{2}}
-a_{\bar{2}}D_{\bar{0}+}^{-1}a_{\bar{3}}D_{\bar{0}+}^{-1}a_{\bar{3}}D_{\bar{0}+}^{-1}a_{\bar{2}}
\\[5pt]
	\{J_{\bar{0}+},J_{\bar{0}+}\}^{[-2]}&=&
-a_{\bar{2}}D_{\bar{0}+}^{-1}a_{\bar{2}}+a_{\bar{3}}D_{\bar{0}+}^{-1}a_{\bar{3}}D_{\bar{0}+}^{-1}a_{\bar{2}}+a_{\bar{2}}D_{\bar{0}+}^{-1}a_{\bar{3}}D_{\bar{0}+}^{-1}a_{\bar{3}}
+\nonumber\\
&& +a_{\bar{3}}D_{\bar{0}+}^{-1}a_{\bar{2}}D_{\bar{0}+}^{-1}a_{\bar{3}}-a_{\bar{3}}D_{\bar{0}+}^{-1}a_{\bar{3}}D_{\bar{0}+}^{-1}a_{\bar{3}}D_{\bar{0}+}^{-1}a_{\bar{3}}
\nonumber
\end{eqnarray}

\subsection{Different ways of presenting the Poisson bivector.}
\label{sec:EquivalenceRelationsForTheta}
In our formalism the Poisson bivector is defined
modulo some equivalence relation. 
This is because we calculate $\{F(J),G(J)\}$ assuming
that $F$ and $G$ are invariant under the gauge transformations 
\begin{equation}\label{GaugeTransformationsXiZero}
\delta_{\xi_{\bar{0}}}J=-D\xi_{\bar{0}}
\end{equation}
This means that when we are calculating $\{F,G\}$, we are
assuming that $F$ and $G$ are such that
\[
D_{\bar{0}+}{\delta F\over \delta J_{\bar{0}+}}+
\left[J_{\bar{2}+}, {\delta F\over \delta J_{\bar{2}+}}\right]+
\left[J_{\bar{3}+}, {\delta F\over \delta J_{\bar{3}+}}\right]=0
\]
and the same is true about $G$.
Therefore the Poisson bivector
$\theta^{ij}$ is defined only up to the equivalence 
\begin{equation}\label{EquivalenceOfThetas}
\theta\simeq \theta + v\wedge \delta_{\xi_{\bar{0}}},\;\;\;
\xi_{\bar{0}}\in {\bf g}_{\bar{0}}
\end{equation}
where $v$ is a vector field on the phase space.
Therefore the Poisson bivector (\ref{BoostInvariant})
can be replaced by:
\begin{eqnarray}
&&	\{J_{\bar{0}+},J_{\bar{0}+}\}^{[0]}=2D_{\bar{0}+}  \label{FirstLine} \\
&&	\{J_{\bar{3}+},J_{\bar{3}+}\}^{[0]}=-2a_{\bar{2}}  \label{SecondLine}
\end{eqnarray}
(The Poisson brackets of the components not listed are zero.)

\subsection{Symplectic leaves}
When we write the Poisson bracket $\{,\}^{[0]}$ in the form
(\ref{FirstLine}), (\ref{SecondLine}), 
it becomes obvious that this Poisson structure
is degenerate. Indeed, the brackets do not involve $J_{\bar{2}}$ at
all. Degenerate Poisson brackets define submanifolds in the
phase space known as ``symplectic leaves''. 
Consider a point $x$ of the phase space. Given a 1-form $\lambda$ at $x$, we
can contract it with the Poisson bivector $\theta(x)$ and
get a vector $\theta^{ij}(x)\lambda_j$ in the tangent space at the point $x$.
Therefore $\theta(x)$ defines a linear map $\theta(x): T^*_xM\to T_xM$.
The image of this map is a subspace  $\mbox{Im}\theta(x)\subset T_xM$. 
It turns out that when we vary $x$, the collection of  spaces
$\mbox{Im}\theta(x)$ is an integrable distribution, in the sense that
there exists a foliation of $M$ by submanifolds $N\subset M$ such that
the tangent space to $N$ at every point coincides with $\mbox{Im}\theta$.
(There is a family of submanifolds $N$, parametrized by codim$N$ 
parameters; they are known as  ``the integral manifolds of the distribution
$\mbox{Im}\theta$''.)
The submanifolds $N$ are called ``symplectic leaves'':
\vspace{10pt}

\begin{centering}
	\hfill
\includegraphics[width=2.0in]{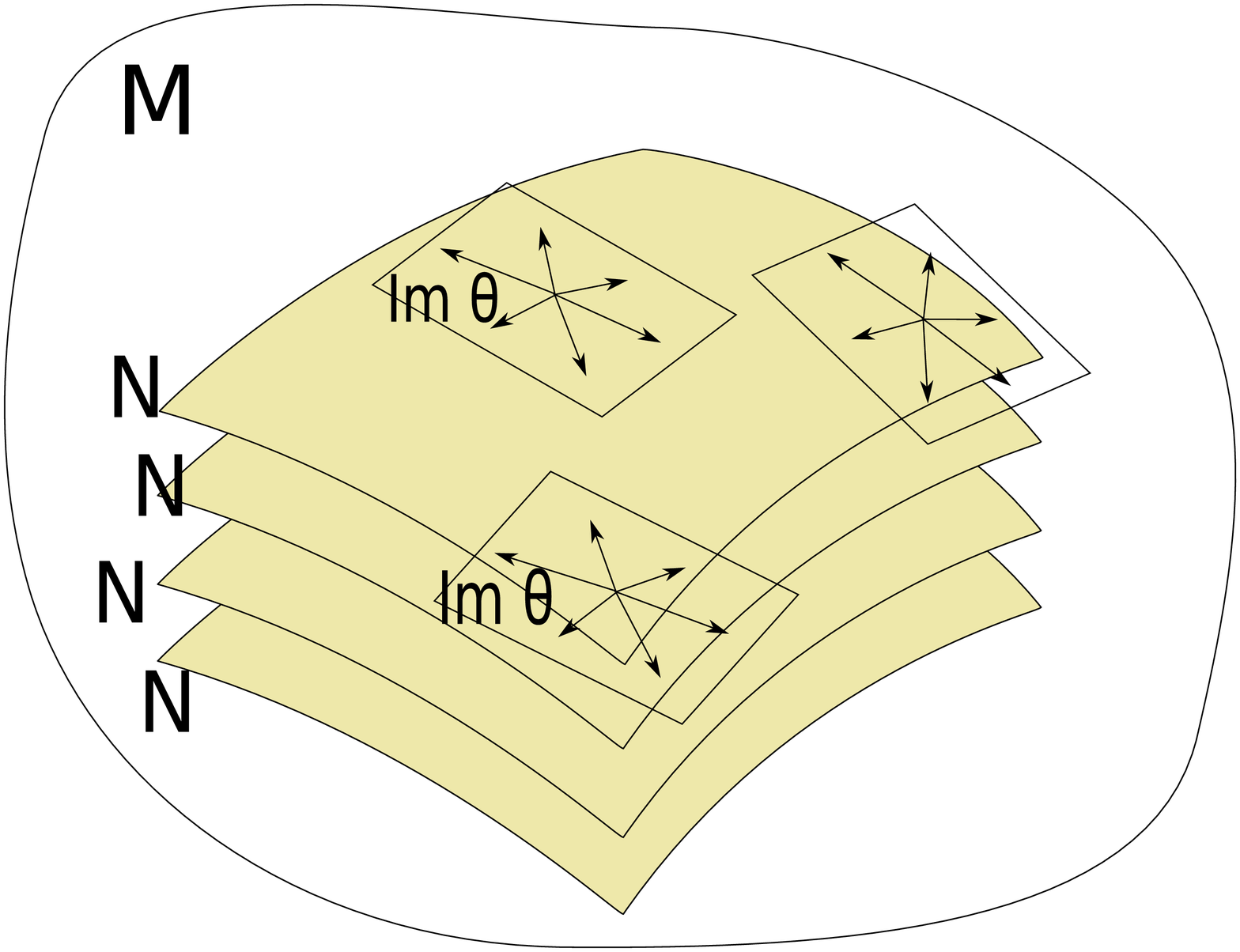}
\hfill
\end{centering}

\vspace{10pt}
\noindent
The symplectic leaves can also be described in terms of the Casimir
functionals. A function $C$ on the phase space is called Casimir
function (or functional),
if its Poisson bracket with any other functional vanishes.
The nondegenerate Poisson brackets do not have any Casimir
functionals, but the degenerate brackets do. 
The symplectic leaves can be characterized as the common level
sets of all the Casimir functionals.

Let us first discuss $\{,\}^{[0]}$ in the bosonic sector, 
Eq. (\ref{FirstLine}). Bosonic degrees of freedom are
$J_{\bar{0}+}$ and $J_{\bar{2}+}$. The symplectic leaves are the manifolds
with the constant eigenvalues of $J_{\bar{2}+}$.
Indeed, the only Casimir functionals are the gauge invariant
functionals of $J_{\bar{2}+}$ only. But the gauge invariance
(\ref{GaugeTransformationsXiZero}) acts on $J_{\bar{2}+}$ by the
conjugation, and therefore the Casimir functions are precisely the
functionals of the eigenvalues of $J_{\bar{2}+}$. Therefore
the symplectic leaves are the level sets of the eigenvalues 
(spectral invariants)
of $J_{\bar{2}+}$. 

The symplectic leaves are compatible with the
Virasoro constraint (\ref{VirasoroConstraintOnJTwo})
in a sense that,
if one point on the symplectic leaf satisfies (\ref{VirasoroConstraintOnJTwo}),
then all the points on this leaf also satisfy (\ref{VirasoroConstraintOnJTwo}).
The bosonic part 
${\bf g}_{ev}= {\bf g}_{\bar{0}} \oplus  {\bf g}_{\bar{2}}$ 
of $psu(2,2|4)$ is $so(2,4)\oplus so(6)$.
It turns out, because of the properties of ${\bf g}=psu(2,2|4)$, that
any $J_{\bar{2}+}$ satisfying (\ref{VirasoroConstraintOnJTwo})
is of the form
\[
J_{\bar{2}+}=
\left(\begin{array}{cccccc}
{ 0}	& \alpha	&  0	&  0 & 0 & 0 \\
{- \alpha}	& 0	& 0	&  0 & 0 & 0 \\
{ 0}	& 0	& 0	&  0 & 0 & 0 \\
{ 0}	& 0	& 0 	&  0 & 0 & 0 \\
{ 0}	& 0	& 0 	&  0 & 0 & 0 \\
{ 0}	& 0	& 0 	&  0 & 0 & 0 
\end{array}\right)_{so(2,4)}\!\!\!\bigoplus \;\;	
\left(\begin{array}{cccccc}
{ 0}	& \alpha	&  0	&  0 & 0 & 0 \\
{ -\alpha}	& 0	& 0	&  0 & 0 & 0 \\
{ 0}	& 0	& 0	&  0 & 0 & 0 \\
{ 0}	& 0	& 0 	&  0 & 0 & 0 \\
{ 0}	& 0	& 0 	&  0 & 0 & 0 \\
{ 0}	& 0	& 0 	&  0 & 0 & 0 
\end{array}\right)_{so(6)}	
\]
up to the conjugation. Here $\alpha=\alpha(\tau^+,\tau^-)$ are
some functions of $\tau^+$ and $\tau^-$. The string worldsheet
fields are defined up to conformal transformations
$(\tau^+,\tau^-)\mapsto (F^+(\tau^+),F^-(\tau^-))$
where $F^+$ and $F^-$ are arbitrary functions (with $(F^+)'\neq 0$,
$(F^-)'\neq 0$).   
Eq. (\ref{MTFirstEqM}) implies that $\partial_-\alpha=0$.
This means that we can do such a conformal transformation
$\tau^+\mapsto F^+(\tau^+)$, or in other words
choose the worldsheet coordinates $(\tau^+,\tau^-)$ in such
a way,  that $\alpha$ is constant and equal to $1$:
\begin{equation}\label{SymplecticLeaf}
\alpha(\tau^+,\tau^-)\equiv 1
\end{equation}
In the following discussion we will use these particular coordinates.
Notice that {\bf Eq. (\ref{SymplecticLeaf})  defines a symplectic
leaf.} The hidden relativistic symmetry of
Section \ref{sec:HiddenRelativisticSymmetry} preserves the condition
(\ref{SymplecticLeaf}).

\subsection{Relation between $\theta^{[0]}$ 
and the chiral WZW bracket }
Now we want to explain that $\theta^{[0]}$ can be thought of as a Hamiltonian
reduction of the Kirillov bracket on the coadjoint orbit of the
Kac-Moody algebra.
The discussion in this subsection is similar to 
\cite{MR2068787}; Eq. (\ref{SymplecticFormLCSuperstring})
easily follows from the results of \cite{Bakas:1995bm}.

Let us first discuss the Poisson bracket of the bosonic
fields
$J_{\bar{0}+}$, and then turn to the fermions.
Let us choose the gauge where $J_{\bar{2}+}$ is equal to:
\begin{equation}\label{SpecialGaugeChoice}
J_{\bar{2}+}=
\left(\begin{array}{cccccc}
{ 0}	& 1	&  0	&  0 & 0 & 0 \\
{ -1}	& 0	& 0	&  0 & 0 & 0 \\
{ 0}	& 0	& 0	&  0 & 0 & 0 \\
{ 0}	& 0	& 0 	&  0 & 0 & 0 \\
{ 0}	& 0	& 0 	&  0 & 0 & 0 \\
{ 0}	& 0	& 0 	&  0 & 0 & 0 
\end{array}\right)_{\!\!\!so(2,4)}\!\!\!\bigoplus \;\;	
\left(\begin{array}{cccccc}
{ 0}	& 1	&  0	&  0 & 0 & 0 \\
{ -1}	& 0	& 0	&  0 & 0 & 0 \\
{ 0}	& 0	& 0	&  0 & 0 & 0 \\
{ 0}	& 0	& 0 	&  0 & 0 & 0 \\
{ 0}	& 0	& 0 	&  0 & 0 & 0 \\
{ 0}	& 0	& 0 	&  0 & 0 & 0 
\end{array}\right)_{\!\!\!so(6)}	
\end{equation}
(Remember that the bosonic part of $psu(2,2|4)$ is 
${\bf g}_{\bar{0}}\oplus{\bf g}_{\bar{2}}=so(2,4)\oplus so(6)$.)
In this gauge $F$ and $G$ become functions
of $J_{\bar{0}+}$, invariant under the {\em residual gauge transformations}:
\begin{equation}\label{ZetaGaugeTransformation}
\delta_{\zeta} J_{\bar{0}+}=D_{\bar{0}+}\zeta,\;\;\;\; \zeta\in{\bf h}
\end{equation}
 where $\zeta$ belongs
to the subalgebra ${\bf h}\subset {\bf g}_{\bar{0}}$
 which stabilizes $J_{\bar{2}+}$:
\[
{\bf h}\;:\;\;\;
\left(
\begin{array}{cccccc}
0 & 0 & 0 & 0 & 0 & 0 \\
0 & 0 & 0 & 0 & 0 & 0 \\
0 & 0 & * & * & * & * \\
0 & 0 & * & * & * & * \\
0 & 0 & * & * & * & * \\
0 & 0 & * & * & * & *
\end{array} 
\right)_{\!\!\!\! so(2,4)}
\!\!\!\!\!\!\!\! \bigoplus
\left(
\begin{array}{cccccc}
0 & 0 & 0 & 0 & 0 & 0 \\
0 & 0 & 0 & 0 & 0 & 0 \\
0 & 0 & * & * & * & * \\
0 & 0 & * & * & * & * \\
0 & 0 & * & * & * & * \\
0 & 0 & * & * & * & *
\end{array} 
\right)_{\!\!\!\! so(6)}
\]
We want to describe the Poisson bracket (\ref{FirstLine})
on the functions invariant under the gauge transformations
(\ref{ZetaGaugeTransformation}). We can look at it in the 
following way. Consider first the bracket (\ref{FirstLine})
on arbitrary functions $F(J_{\bar{0}+})$, not necessarily
gauge invariant. This is essentially the 
chiral WZW bracket \cite{Witten:1983ar},
or equivalently the Kirillov bracket on the coadjoint
orbit of the Kac-Moody algebra. Let us parametrize the currents
$J_{\bar{0}+}$ in terms of the monodromy $f$:
\[
J_{\bar{0}+}=-\partial_+ f f^{-1}
\]
The symplectic structure corresponding to (\ref{FirstLine}) is 
\begin{equation}\label{SymplecticWZW}
\Omega_{WZW}=
\int d\tau^+ \mbox{tr}\; \left(\delta f f^{-1} \delta(\partial_+f f^{-1})
\right)
\end{equation}
Now we want to describe the bracket (\ref{FirstLine}) 
specifically on the functions
invariant under the gauge transformations (\ref{ZetaGaugeTransformation}). 
Instead of talking about the gauge invariant functions we can consider
functions on a submanifold in the phase space (the "gauge slice")
which is a symplectic complement 
of the orbits of (\ref{ZetaGaugeTransformation}), with respect
to the symplectic form (\ref{SymplecticWZW}). ``Symplectic complement''
means that for any vector $\xi$ tangent to the gauge
slice and any $\zeta\in {\bf h}$ we should have
\begin{equation}\label{SymplecticComplement}
\Omega_{WZW}(\delta_{\xi},\delta_{\zeta})=0
\end{equation}
We choose the following gauge slice:
\begin{equation}\label{NormalFrameGauge}
J_{\bar{0}+}\!= \!\!
\left(
\begin{array}{ccccc}
0		& q^0_+	& q^1_+	& q^2_+	& q^3_+	\\
\tilde{q}^0_+		& 0	& 0	& 0	& 0	\\
-\tilde{q}^1_+		& 0	& 0	& 0	& 0	\\
-\tilde{q}^2_+		& 0	& 0	& 0	& 0	\\
-\tilde{q}^3_+		& 0	& 0	& 0	& 0	
\end{array}
\right)_{\!\!\!\!\! so(1,4)}\!\!\!\!\!\!\!\!\!\bigoplus
\left(
\begin{array}{ccccc}
0		& q^1_+	& q^2_+	& q^3_+	& q^4_+	\\
-q^1_+		& 0	& 0	& 0	& 0	\\
-q^2_+		& 0	& 0	& 0	& 0	\\
-q^3_+		& 0	& 0	& 0	& 0	\\
-q^4_+		& 0	& 0	& 0	& 0	
\end{array}
\right)_{\!\!\!\!\! so(5)}
\end{equation}
This gauge slice satisfies Eq. (\ref{SymplecticComplement}).
This implies that the symplectic structure in terms of the variables
$q^i_+$ is given by the restriction of the symplectic form 
(\ref{SymplecticWZW}) to the subspace of the phase space specified
by the constraint that $\partial_+ f f^{-1}=-J_{\bar{0}+}$ is of the form
(\ref{NormalFrameGauge}).

The gauge choice (\ref{NormalFrameGauge}) was used in 
\cite{MR1102831} --- \cite{Anco:2005b}.  
Geometrically it corresponds to the 
so-called ``normal frame''. The normal frame is the basis in the
normal bundle to the curve such that the covariant derivative of any
element of this basis along the curve is parallel to the tangent
vector to the curve
\cite{Bishop}.

Now let us consider the fermionic part. 
The Poisson bracket 
\[
\{J_{\bar{3}+},J_{\bar{3}+}\}=a_{\bar{2}}
\]
is degenerate. It has symplectic leaves which are described by the
equation
\begin{equation}
	J_{\bar{3}}-J_{\bar{3}}^{(0)}=[J_{\bar{2}+},K_{\bar{1}}]
\end{equation}
where $K_{\bar{1}}$ runs over ${\bf g}_{\bar{1}}$.
The gauge (\ref{SpecialGaugeChoiceFermions}) corresponds to the symplectic
leaf with $J_{\bar{3}}^{(0)}=0$, in other words
\begin{equation}
	J_{\bar{3}}\in 
	\mbox{Im}( \mbox{ad}_{J_{\bar{2}+}}:{\bf g}_{\bar{1}}\rightarrow
{\bf g}_{\bar{3}})
\end{equation}
On this symplectic leaf we have $J_{\bar{3}+}=[J_{\bar{2}+},K_{\bar{1}}]$
and the symplectic form is:
\begin{equation}\label{SymplecticFormLCSuperstring}
\int d\tau^+ \left(\mbox{tr}\; \left(\delta f f^{-1} \delta(\partial_+f f^{-1})
\right) +\mbox{tr}\;\left(\delta K_{\bar{1}} [J_{\bar{2}+},
\delta K_{\bar{1}}]\right)\right)
\end{equation}
This is obviously a closed form\footnote{We have to remember that
$J_{\bar{2}+}$ is gauge fixed to be equal to (\ref{SpecialGaugeChoice}).}, 
and therefore {\bf $\{,\}^{[0]}$ satisfies
the Jacobi identity.}
The bosonic part of the symplectic form (\ref{SymplecticFormLCSuperstring}) 
follows from the action
of the generalized sine-Gordon model which was suggested 
in \cite{Bakas:1995bm}. The possible 
relation with the WZW model on the quantum level
was discussed in \cite{Polyakov:2005ss}.

\subsection{Relation to the results of \cite{Mikhailov:2005sy}.}
If we apply this formailsm to ${\bf g}=so(3)$ we get ${\bf g}_{\bar{0}}=so(2)$
and ${\bf g}_{\bar{2}}$ is the vector representation of $so(2)$. We can write:
\begin{equation}
J_{\bar{2}+}=\left[\begin{array}{c} r\cos\varphi \\ r\sin\varphi \end{array}\right]
\end{equation}
The Poisson bracket $\theta^{[2]}$ at $r=1$ becomes:
\begin{equation}
\{F,G\}^{[2]}  =\int d\tau^+\left(
  {\delta F\over\delta \varphi}\partial_+ {\delta G\over\delta\varphi}  + 
{\delta F\over \delta r}\partial_+ {\delta G\over\delta r} -
\partial_+\varphi \left({\delta F\over\delta r}{\delta G\over\delta\varphi} - 
{\delta F\over\delta \varphi}{\delta G\over\delta r}\right)\right)
\end{equation}
This means that if $F$ depends only on $\varphi$ we get 
$\{r(\tau_1^+),F\}^{[2]}={\delta F\over\delta\varphi(\tau_1^+)}\partial_+\varphi(\tau_1^+)$.
If we consider $F$ as a function of $q_+=\partial_+\varphi$, then
${\delta F\over\delta \varphi}=-\partial_+{\delta F\over\delta q_+}$.
If we fix the Virasoro constraint $r=1$, the Dirac bracket becomes
\[
\{F,G\}_D=\{F,G\}-\{F,r\}\{r,r\}^{-1}\{r,G\}=
\{F,G\}-\int d\tau^+ {\delta F\over\delta q_+}\partial_+q_+\partial_+^{-1}
q_+\partial_+ {\delta G\over\delta q_+}
\]
This agrees with $\theta_1$ of \cite{Mikhailov:2005sy}. We see that resolving
the Virasoro constraint gives an additional nonlocal piece in the Dirac bracket
\cite{Ferapontov}.

\subsection{The relation between $\theta^{[2]}$, $\theta^{[0]}$ and 
$\theta^{[-2]}$}
We have seen that the boost-invariant Poisson bracket (\ref{BoostInvariant})
can be written in the equivalent form (\ref{FirstLine}), (\ref{SecondLine}).
Alternatively, it can be also (using the same equivalence relations,
described in Section \ref{sec:EquivalenceRelationsForTheta}) 
written in the following form:
\begin{eqnarray}\label{FixedJ0Gauge}
\{J_{\bar{2}+},J_{\bar{2}+}\}^{[0]}&=&
2a_{\bar{2}}D_{\bar{0}+}^{-1}a_{\bar{2}} \nonumber \\
\{J_{\bar{2}+},J_{\bar{3}+}\}^{[0]}&=&
2a_{\bar{2}}D_{\bar{0}+}^{-1}a_{\bar{3}} \label{BoostInvariantThroughJ2J3}\\
\{J_{\bar{3}+},J_{\bar{3}+}\}^{[0]}&=&-2 a_{\bar{2}} + 
2a_{\bar{3}}D_{\bar{0}+}^{-1}a_{\bar{3}} \nonumber
\end{eqnarray}
The brackets of the components not listed are zero. In particular, $J_{0+}$ 
commutes with everything. This means that Eq. (\ref{FixedJ0Gauge})  gives
the boost-invariant Poisson bracket in the gauge $J_{0+}=0$.
More explicitly:
\[
	\{F,G\}^{[0]}=
	2 \int d\tau^+ 
\left[ {\delta F\over\delta J_{\bar{2}+}}\; , \;
{\delta F\over\delta J_{\bar{3}+}}\right]
\left[
\begin{array}{cc}
	a_{\bar{2}}D_{\bar{0}+}^{-1}a_{\bar{2}}	&
	a_{\bar{2}}D_{\bar{0}+}^{-1}a_{\bar{3}} 		\\
	a_{\bar{3}}D_{\bar{0}+}^{-1}a_{\bar{2}}			&
	- a_{\bar{2}} + 
	a_{\bar{3}}D_{\bar{0}+}^{-1}a_{\bar{3}}
\end{array}
\right]
\left[
\begin{array}{c} 	
	\delta G\over \delta J_{\bar{2}+} \\
	\delta G\over \delta J_{\bar{3}+} 
\end{array}
\right]
\]
In other words, $\theta^{[0]}$ in this ``picture'' is:
\begin{equation}\label{BoostInvariantExplicitly}
\theta^{[0]}=
\left[
\begin{array}{cc}
	2a_{\bar{2}}D_{\bar{0}+}^{-1}a_{\bar{2}}	&
	2a_{\bar{2}}D_{\bar{0}+}^{-1}a_{\bar{3}} 		\\
	2a_{\bar{3}}D_{\bar{0}+}^{-1}a_{\bar{2}}			&
	- 2a_{\bar{2}} + 
	2a_{\bar{3}}D_{\bar{0}+}^{-1}a_{\bar{3}}
\end{array}
\right]
\end{equation}
Let us also explicitly write $\theta^{[2]}$ in the same gauge:
\[
\{F,G\}^{[2]}=\int d\tau^+
\left[ {\delta F\over\delta J_{\bar{2}+}}\; , \;
{\delta F\over\delta J_{\bar{3}+}}\right]
\left[
\begin{array}{cc}
	-D_{\bar{0}+}+a_{\bar{3}}a_{\bar{2}}^{-1}a_{\bar{3}} 	&
	a_{\bar{3}}a_{\bar{2}}^{-1}D_{\bar{0}+} 			\\
	D_{\bar{0}+}a_{\bar{2}}^{-1}a_{\bar{3}}			&
	D_{\bar{0}+}a_{\bar{2}}^{-1} D_{\bar{0}+}
\end{array}
\right]
\left[
\begin{array}{c} 	
	\delta G\over \delta J_{\bar{2}+} \\
	\delta G\over \delta J_{\bar{3}+} 
\end{array}
\right]
\]
The corresponding symplectic structure $\Omega^{[2]}=(\theta^{[2]})^{-1}$ is:
\begin{eqnarray}
&&
\Omega^{[2]}=(\theta^{[2]})^{-1}= \label{SymplecticStructureHighestGrade}\\
&&=\int d\tau^+
\left[ {\delta J_{\bar{2}+}}\; , \;
{\delta J_{\bar{3}+}}\right] D_{\bar{0}+}^{-1}
\left[\begin{array}{cc}
	-D_{\bar{0}+}	& a_{\bar{3}}	 				\\
	a_{\bar{3}}	& a_{\bar{2}}-a_{\bar{3}}D_{\bar{0}+}^{-1}a_{\bar{3}}
\end{array}\right]
D_{\bar{0}+}^{-1}
\left[
\begin{array}{c}
	\delta J_{\bar{2}+} \\ \delta J_{\bar{3}+}
\end{array}
\right]
\nonumber
\end{eqnarray}
Now let us bring $\theta^{[-2]}$ to the same gauge. 
Section \ref{sec:EquivalenceRelationsForTheta} allows us to present 
$\{,\}^{[-2]}$ as follows:
\begin{equation}\label{RewritingThetaMinusTwo}
	\{F,G\}^{[-2]}=\int d\tau^+
\left[ {\delta F\over\delta J_{\bar{2}+}}\; , \;
{\delta F\over\delta J_{\bar{3}+}}\right]
\left[\begin{array}{cc}
	\theta^{[-2]}_{\bar{2}\bar{2}}
	& \theta^{[-2]}_{\bar{2}\bar{3}} \\[5pt]
	\theta^{[-2]}_{\bar{3}\bar{2}} 
	& \theta^{[-2]}_{\bar{3}\bar{3}}
\end{array}\right]
\left[
\begin{array}{c}
	\delta G/ \delta J_{\bar{2}+} \\ \delta G/ \delta J_{\bar{3}+}
\end{array}
\right]
\nonumber
\end{equation}
where
\begin{eqnarray}
\theta^{[-2]}_{\bar{2}\bar{2}}
	&=&-{\cal O}_{2222}+{\cal O}_{23322}+
	{\cal O}_{22332}+{\cal O}_{23232}-{\cal O}_{233332} \nonumber\\[5pt]
\theta^{[-2]}_{\bar{2}\bar{3}}
	&=&-{\cal O}_{2223}-{\cal O}_{2232}-{\cal O}_{2322}+\\
	&&+{\cal O}_{23332}+{\cal O}_{23323}+{\cal O}_{23233}+{\cal O}_{22333}
	-{\cal O}_{233333}
	\nonumber\\[5pt]
\theta^{[-2]}_{\bar{3}\bar{3}}
	&=&
	{\cal O}_{222} - {\cal O}_{2233}-{\cal O}_{2323}
	-{\cal O}_{3232} -{\cal O}_{3322}-{\cal O}_{2332}-{\cal O}_{3223}
	+\nonumber \\
&&	+{\cal O}_{23333}+{\cal O}_{32333}+{\cal O}_{33233}
	+{\cal O}_{33323}+{\cal O}_{33332}-\nonumber \\
&&	-{\cal O}_{333333}\nonumber
\end{eqnarray}
Here we introduced the notations ${\cal O}_{\bar{j}_1\bar{j}_2\ldots\bar{j}_n}$:
\begin{equation}
	{\cal O}_{ \bar{j}_1 \bar{j}_2 \bar{j}_3\ldots \bar{j}_n}=
	a_{\bar{j}_1}D^{-1}_{\bar{0}+}a_{\bar{j}_2}D^{-1}_{\bar{0}+}a_{\bar{j}_3}D^{-1}_{\bar{0}+}
	\cdots D^{-1}_{\bar{0}+}a_{\bar{j}_n}
\end{equation}
The indices  of ${\cal O}$ run over $\bar{j}\in\{\bar{2},\bar{3}\}$.
An explicit computation using (\ref{SymplecticStructureHighestGrade})
and (\ref{BoostInvariantExplicitly}) and (\ref{RewritingThetaMinusTwo})
shows that
\begin{equation}\label{RelationBetweenBrackets}
4\theta^{[-2]}= \theta^{[0]} (\theta^{[2]})^{-1} \theta^{[0]}
\end{equation}
We see that the Poisson bivector has the same structure as
in the case of the string in ${\bf R}\times S^2$ considered in 
\cite{Mikhailov:2005sy}:
\begin{equation}\label{CanonicalPoissonStructure}
	\theta^{can}=\theta^{[2]}+\theta^{[0]}
	+{1\over 4}\theta^{[0]} (\theta^{[2]})^{-1} \theta^{[0]}
\end{equation}

\subsection{Poisson bracket and the monodromy matrix}
The monodromy matrix can be defined as a path ordered exponential:
\begin{equation}\label{MonodromyMatrix}
M(z)^{ab}=\left[P\exp \int_{-\infty}^{\infty}
 d\tau^+ \left( -z J_{1+} - J_{0+} - {1\over z} J_{3+} - {1\over z^2} J_{2+}
\right)\right]^{ab}
\end{equation}
Although we are using the gauge $J_{1+}=0$ we have included $J_{1+}$
in the definition of $M$ for the convenience of notations; we will consider
the variational derivative $\delta M\over \delta J_{1+}$ at $J_{1+}=0$.
The monodromy matrix (\ref{MonodromyMatrix}) is a functional of the currents, 
and we can calculate its Poisson
brackets with the other functionals. Let us first study the properties
of the variational derivatives $\delta M\over\delta J_+(\tau^+)$.
We will define $\delta M\over\delta J_+(\tau^+)$ as a matrix such that
\begin{equation}
\int_{-\infty}^{+\infty} d\tau^+\; \mbox{str}\; 
\delta J_+ {\delta M^{ab}\over\delta J_+(\tau^+)} = 
\delta M^{ab}
\end{equation}
Consider the following identities:
\begin{eqnarray}
D_{0+} {\delta M^{ab}\over \delta J_{3+}} + 
\left[ J_{3+}, {\delta M^{ab}\over \delta J_{2+}} \right] +
{1\over z^4} \left[J_{2+}, {\delta M^{ab}\over \delta J_{1+}}\right] = 0 
\label{FirstIdentityForM}
\\
 D_{0+} {\delta M^{ab}\over \delta J_{2+} } 
+ {1\over z^4} \left[ J_{3+}, {\delta M^{ab}\over \delta J_{1+} } \right]
+ {1\over z^4} \left[ J_{2+}, {\delta M^{ab}\over \delta J_{0+} } \right] =  0
\label{SecondIdentityForM}
\\
D_{0+}{\delta M^{ab}\over \delta J_{1+}} + 
\left[ J_{2+}, {\delta M^{ab}\over \delta J_{3+}}\right] +
\left[ J_{3+}, {\delta M^{ab}\over \delta J_{0+}}\right]=0
\label{ThirdIdentityForM}
\\
D_{0+}{\delta M^{ab}\over \delta J_{0+}} +
\left[ J_{2+}, {\delta M^{ab}\over \delta J_{2+}}\right] +
\left[ J_{3+}, {\delta M^{ab}\over \delta J_{3+}}\right]=0
\label{FourthIdentityForM}
\end{eqnarray}
which follow from the invariance of  (\ref{MonodromyMatrix}) under the gauge
transformations  with $z$-dependent parameters; for example 
(\ref{FirstIdentityForM}) follows from the formula:
\[
\delta_{\alpha_3} 
\left(D_{0+}+zJ_{1+}+{1\over z}J_{3+}+{1\over z^2}J_{2+} \right) =
\left[ D_{0+}+zJ_{1+}+{1\over z}J_{3+}+{1\over z^2}J_{2+} \;,\; 
{1\over z}\alpha_3 \right]
\]
which should be understood as $\delta_{\alpha_3}J_{3+}=D_{0+}\alpha_3$
and $\delta_{\alpha_3}J_{1+}={1\over z^4}[J_{2+},\alpha_3]$.
Notice that (\ref{FourthIdentityForM}) is the statement of 
${\bf g}_{\bar{0}}$ gauge invariance. 
Eqs. (\ref{FirstIdentityForM}), (\ref{SecondIdentityForM}) and (\ref{ThetaTwo}) imply:
\begin{eqnarray}
&& \{J_{2+}, M^{ab} \}^{[2]} = 
{1\over z^4}  \left[ J_{2+}, {\delta M^{ab}\over \delta J_{0+}} \right] 
\\
&& \{J_{3+}, M^{ab} \}^{[2]} = 
-{1\over z^4} D_{0+}{\delta M^{ab}\over \delta J_{1+}}
\end{eqnarray}
On the other hand Eqs. (\ref{BoostInvariantThroughJ2J3}) imply:
\begin{eqnarray}
&& \{J_{2+}, M^{ab} \}^{[0]} = 
-2\left[ J_{2+}, {\delta M^{ab}\over \delta J_{0+}} \right] \\
&& \{J_{3+}, M^{ab} \}^{[0]} = 
2D_{0+} {\delta M^{ab}\over \delta J_{1+}}
\end{eqnarray}
These equations together with (\ref{SecondIdentityForM}) imply that 
\begin{eqnarray}& M(z)^{ab} \mbox{  is a Casimir functional of the Poisson bracket  } &
\nonumber \\ 
& \{,\}^{[0]}+2z^4\{,\}^{[2]} &
\label{CasimirFunctional}
\end{eqnarray}
Notice that for $z=1$ Eqs. (\ref{CasimirFunctional}) and 
(\ref{CanonicalPoissonStructure}) imply that $M(1)^{ab}$ is a Casimir
functional of the canonical Poisson bracket $\theta^{str}$.
This is because at $z=1$ the transfer matrix $P\exp\int_C dg g^{-1}$
over the contour $C$
is expressed in terms of the string worldsheet fields at the 
endpoints of $C$. 

This reasoning is rather formal, because we have not taken into
account the boundary terms. We have not studied the boundary conditions in this paper.
But we expect that with
 the boundary conditions properly taken into account,
only the traces $\mbox{tr}M(z)$ are Casimir functionals. This would imply
that the traces of the monodromy matrix are in involution with respect
to all the compatible Poisson brackets, see the last page of \cite{FaddeevTakhtajan}.

\section{Geometrical meaning of the boost-invariant Poisson bracket}
\label{sec:Geometrical}
In this section we will give a geometrical description of the boost-invariant
Poisson bracket in the special case when the motion of the string is restricted
to ${\bf R}\times S^N$. 

The classical string is equivalent to the nonlinear
sigma-model (\ref{NLSMequation}) with the Virasoro 
constraints (\ref{NLSMwithVirasoroConstraints}). 
As we discussed in Section \ref{sec:LightconeApproach}, through
each point of the string worldsheet pass two lightlike curves $C^+$
and $C^-$ called characteristics. In this section we will consider
the projections of $C^+$ and $C^-$ to $S^5$, and also denote them
$C^+$ and $C^-$. The discussion in this section applies to $S^N$
for arbitrary $N$.

Let $K^+$ and $K^-$ be the unit vectors on the worldsheet orthogonal
to $C^+$ and $C^-$ respectively.

\vspace{9pt}

\begin{centering}
\begin{minipage}[t]{0.5\linewidth}
	\begin{centering}
	\hfill
	{
\includegraphics[width=1.5in]{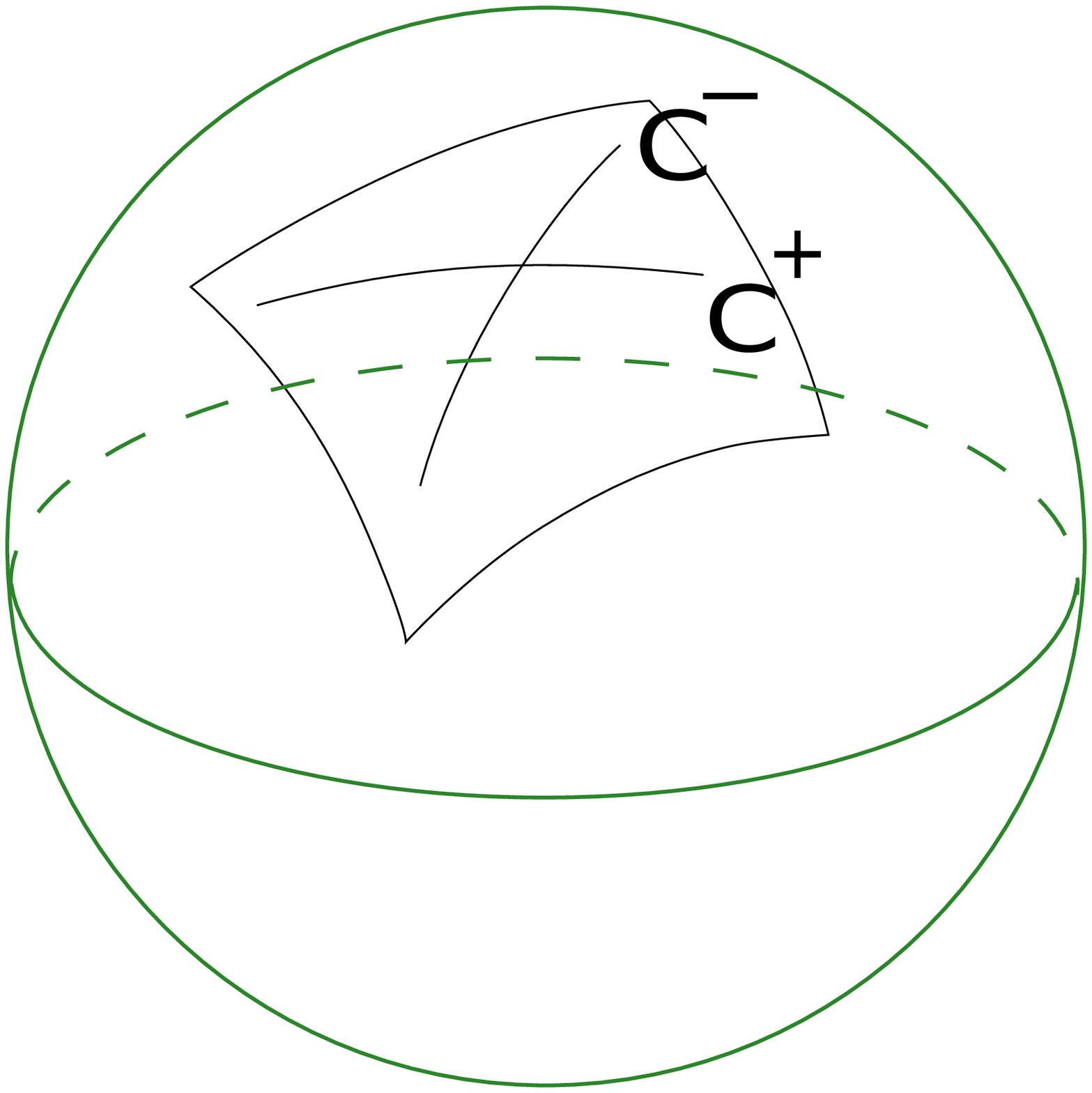}
}
\hfill
\end{centering}
\end{minipage}
\begin{minipage}[t]{0.5\linewidth}
	\begin{centering}
	\hfill
	{
\includegraphics[width=2.0in]{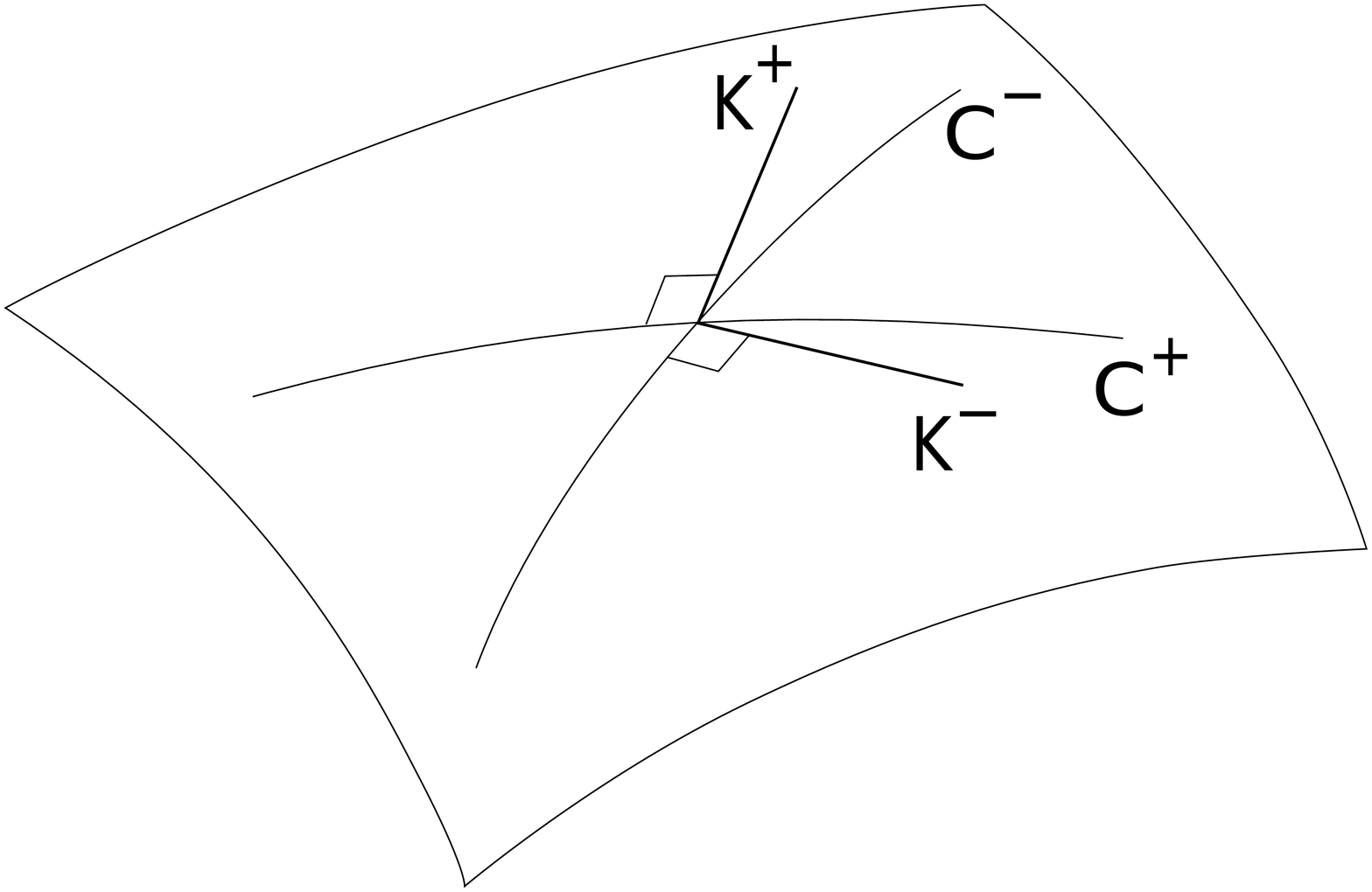}
}
\hfill
\end{centering}
\end{minipage}
\end{centering}

\noindent
Consider also the normal bundle ${\cal N}$ to $\Sigma$ in $S^N$.
It consists of those vectors in $TS^N$ which are orthogonal to $T\Sigma$.
The rank of ${\cal N}$ is $N-2$, for example for $S^N=S^5$ we
get three normal vectors at each point of $\Sigma$.
Let us consider the vector bundles ${\cal N}\oplus K^+$
and ${\cal N}\oplus K^-$, both having the rank $N-1$.

\vspace{10pt}

\begin{centering}
\begin{minipage}[t]{0.33\linewidth}
	{
\includegraphics[width=1.6in]{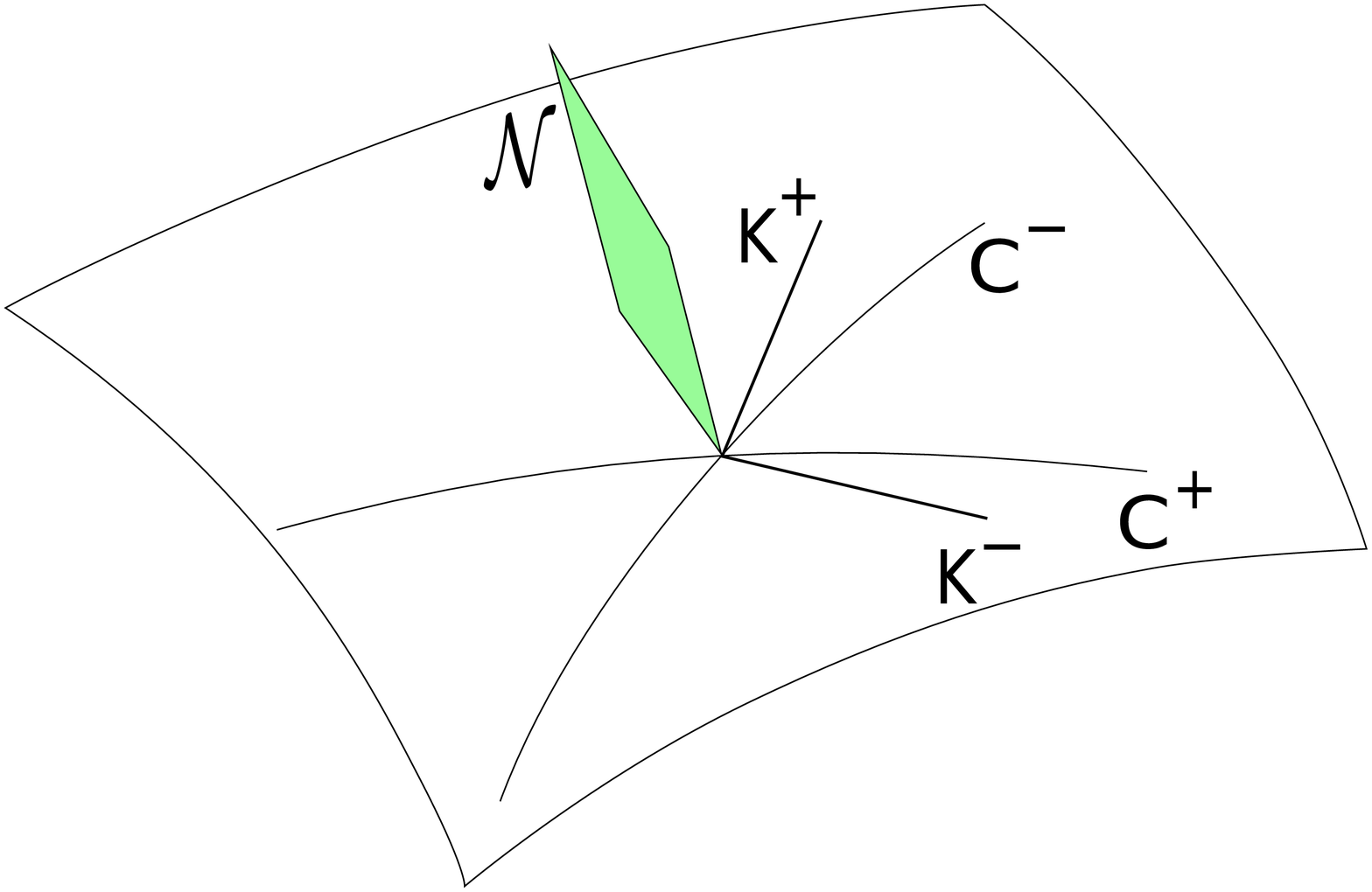}
}
\end{minipage}
\begin{minipage}[t]{0.33\linewidth}
	{
\includegraphics[width=1.6in]{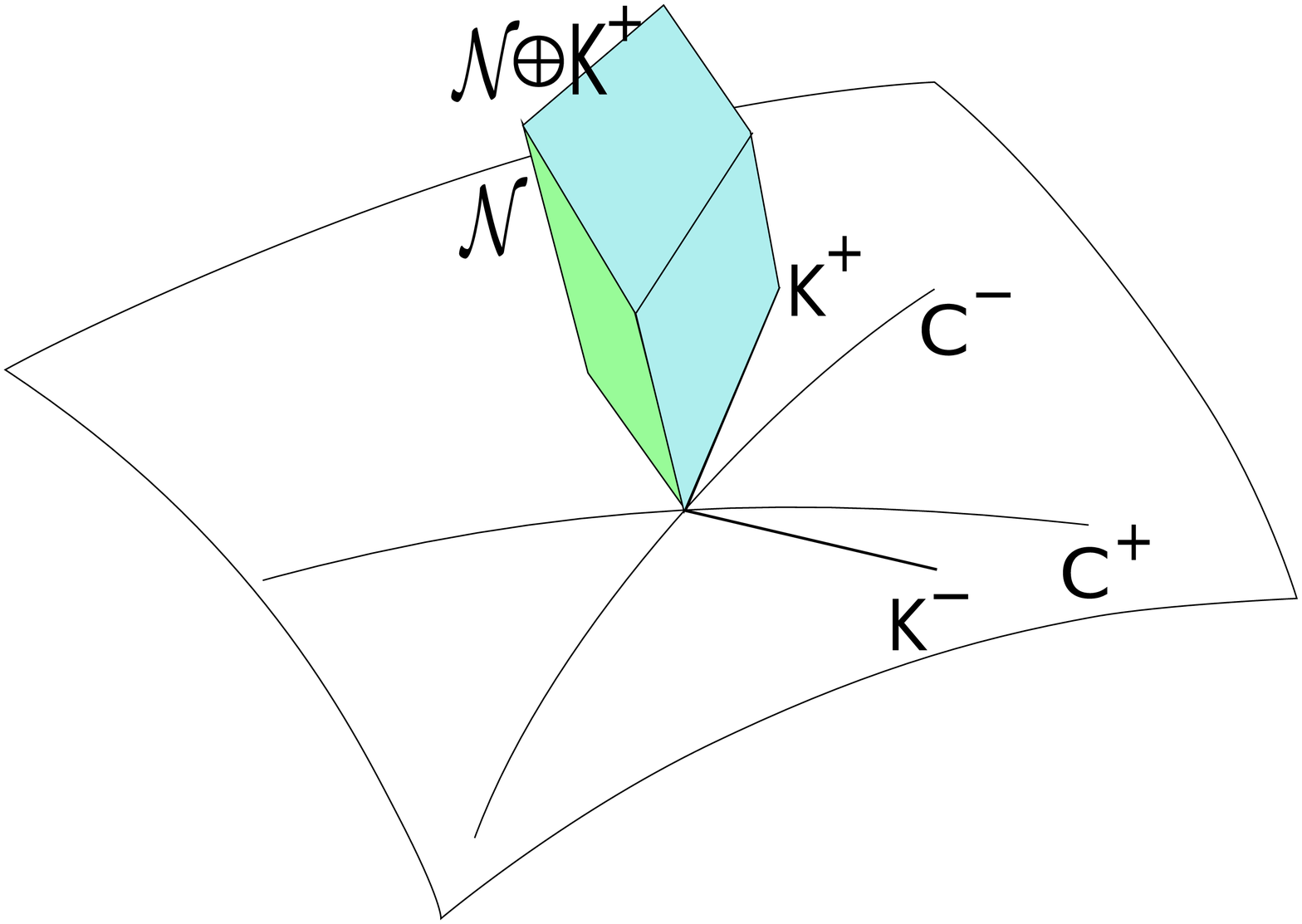}
}
\end{minipage}
\begin{minipage}[t]{0.33\linewidth}
	{
\includegraphics[width=1.6in]{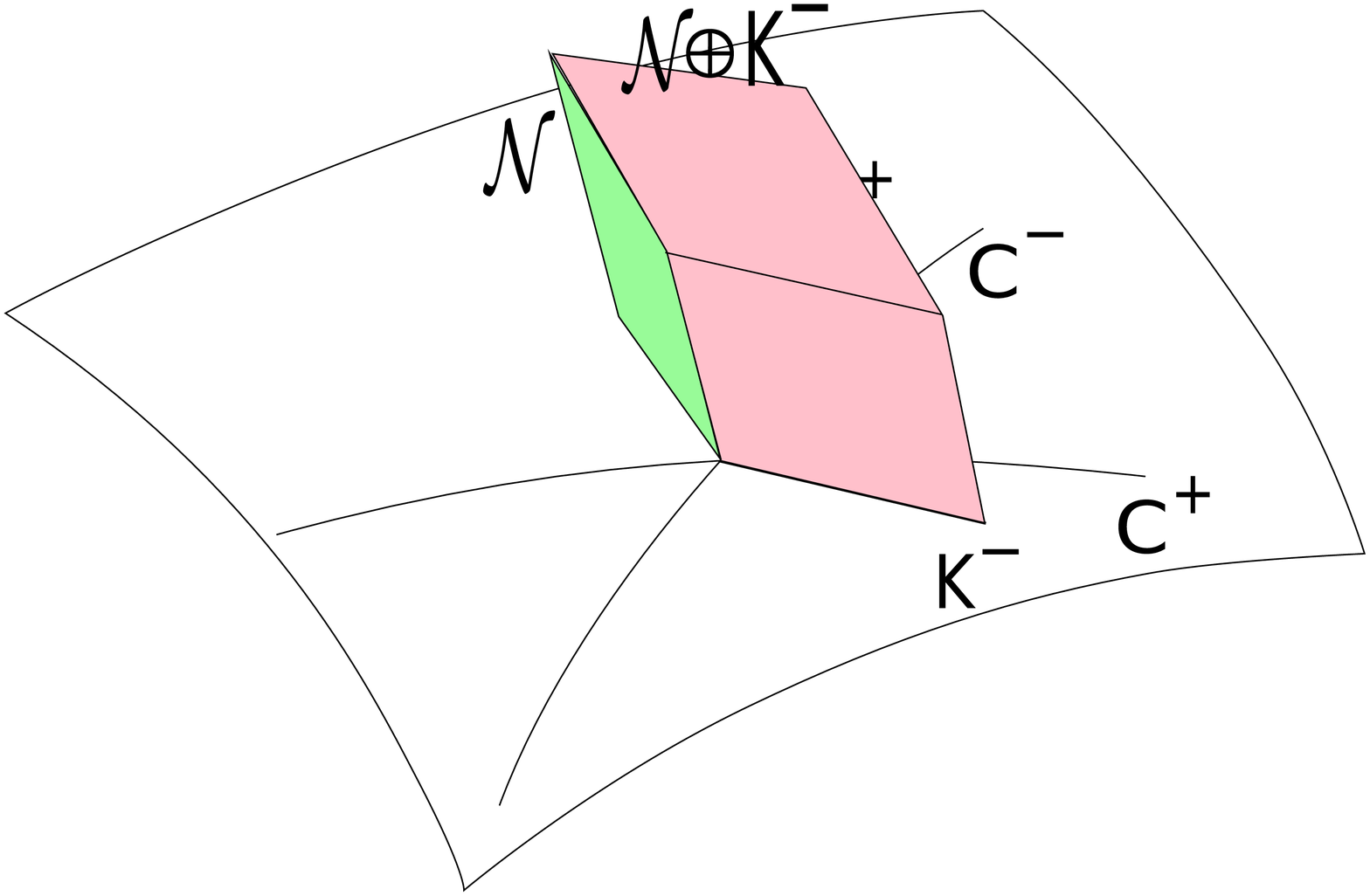}
}
\end{minipage}
\end{centering}

\noindent
Let us restrict the standard Levi-Civita connection to ${\cal N}\oplus K^+$
and ${\cal N}\oplus K^-$. We will denote the restricted connections $\nabla^L$
and $\nabla^R$:
\begin{eqnarray}
	&&	\nabla^L=D_{\bar{0}}|_{ {\cal N}\oplus K^+}\nonumber \\
	&&	\nabla^R=D_{\bar{0}}|_{ {\cal N}\oplus K^-}
\end{eqnarray}
``Restricted connection'' means that, for example, 
\[
\nabla^L v ={\cal P}_{ {\cal N}\oplus K^+ } D_{\bar{0} }\; v
\]
where $v$ is a section of ${\cal N}\oplus K^+$ and 
${\cal P}_{ {\cal N}\oplus K^+ }$ is the projection on 
${\cal N}\oplus K^+\subset TS^N$.
It is easy to verify that both $\nabla^L$ and $\nabla^R$ are flat connections:
\begin{equation}
[\nabla_+^L,\nabla_-^L]=0,\;\;\;
[\nabla_+^R,\nabla_-^R]=0
\end{equation}
This follows from the string worldsheet equations of motion.

Let us introduce some trivialization of ${\cal N}$.
A trivialization is a choice of $N-2$ sections 
${\bf e}_1,\ldots {\bf e}_{N-2}$ of ${\cal N}$ which form an orthonormal
system:
\[
({\bf e}_i,{\bf e}_j)=\delta_{ij}
\]
Notice that the trivialization of ${\cal N}$ defines the trivializations of 
both ${\cal N}\oplus K^+$ and ${\cal N}\oplus K^-$.  
Indeed, to get an orthonormal system in ${\cal N}\oplus K^+$
we just add to ${\bf e}_1,\ldots {\bf e}_{N-2}$ the unit vector
in $K^+$. 

Having the trivializations ${\cal N}\oplus K^+\simeq {\bf R}^{N-1}$
and ${\cal N}\oplus K^-\simeq {\bf R}^{N-1}$ 
we can consider the monodromies of $\nabla^L$
and $\nabla^R$. The monodromies are  the orthogonal
matrices $g^L$ and $g^R$ satisfying
the equations:
\begin{equation}\label{EquationForGLandGR}
\nabla^L g^L=0,\;\;\;\nabla^R g^R=0
\end{equation}
With these notations, the boost-invariant symplectic structure is given by
the following formula:
\begin{eqnarray}\label{BoostInvariantGeometric}
\Omega&=&\oint\left[8\delta \varphi *d\delta\varphi +\right.\label{Omega}\\
&&\left.+\mbox{tr}\left((\delta g_L g_L^{-1}) \delta (d g_L g_L^{-1})\right)-
\mbox{tr}\left((\delta g_R g_R^{-1}) \delta (d g_R g_R^{-1})\right)
\right]
\nonumber
\end{eqnarray}
where $2\varphi$ is the angle between $C^+$ and $C^-$.
One can verify by an explicit calculation 
that (\ref{Omega}) does not depend on the choice of the contour, and
on the choice of the trivialization of ${\cal N}$.
To prove that Eq. (\ref{Omega})
is equivalent to 
Eq. (\ref{SymplecticWZW}) we notice that 
$f$ is related to $g_L$ and $g_R$ in the
following way:
\begin{equation}\label{GvsGLGR}
f=	\left[	
		\begin{array}{cc} 1	 & {\bf 0} \\ 
				{\bf 0}	 & g_R^{-1}
		\end{array}
	\right]
\left[	\begin{array}{ccc}	\cos 2\varphi & -\sin 2\varphi 	&{\bf 0}\\
				\sin 2\varphi & \cos 2\varphi	&{\bf 0}\\
				{\bf 0}	      & {\bf 0}		&{\bf 1}\\
	\end{array}
\right]
	\left[
		\begin{array}{cc} 1	& {\bf 0} \\
				{\bf 0}	& g_L
		\end{array}
	\right]
\end{equation}

\vspace{10pt}
{\small
\noindent Let us first choose a gauge so that
$J_{\bar{2}-}=\left(\begin{array}{lll}
{ 0}	& 1	&  0_{1\times (N-1)}\\
{ -1}	& 0	&  0_{1\times (N-1)} \\
{ 0_{(N-1)\times 1}}	& 0_{(N-1)\times 1}	&  0_{(N-1)\times (N-1)}
\end{array}\right)$.
In this gauge $J_{\bar{0}+}=
\left(\begin{array}{ll}
	{ 0_{2\times 2}}	& 0_{2\times (N-1)}	\\
	{ 0_{(N-1)\times 2} }	& -\partial_+ g_R g_R^{-1}	 
\end{array}\right)$ because $D_{\bar{0}+}J_{\bar{2}-}=0$.
Now let us switch to the gauge where 
$J_{\bar{2}+}=\left(\begin{array}{lll}
{ 0}	& 1	&  0_{1\times (N-1)}\\
{ -1}	& 0	&  0_{1\times (N-1)} \\
{ 0_{(N-1)\times 1}}	& 0_{(N-1)\times 1}	&  0_{(N-1)\times (N-1)}
\end{array}\right)$. This requires the gauge transformation 
$\left(\begin{array}{llll}
1 & 0 & 0 & 0_{1\times (N-2)} \\
0 & \cos 2\varphi & -\sin 2\varphi & 0_{1\times (N-2)} \\
0 & \sin 2\varphi & \cos 2\varphi & 0_{1\times (N-2)} \\
0 & 0_{(N-2)\times 1} & 0_{(N-2)\times 1} & {\bf 1}_{(N-2)\times (N-2)} 
\end{array}
\right)$. Finally,
the gauge transformation 
$\left(\begin{array}{ll}
{ 1_{2\times 2} }	& 0_{2\times (N-1)} \\
{ 0_{(N-1)\times 2} }	& g_L 
\end{array}\right)$
brings us to the normal frame. 
}
\vspace{10pt}

\noindent
Eq. (\ref{GvsGLGR}) allows us to prove
Eq. (\ref{Omega}) using the Polyakov-Wiegmann type of identities.
Let us denote:
\[
f_R^{-1}=	\left[	
	\begin{array}{cc} 1	 & {\bf 0} \\ 
			{\bf 0}	 & g_R^{-1}
	\end{array}
		\right]\; , \;\;\;\;
f_{2\varphi}^{-1}=	\left[
\begin{array}{ccc}	\cos 2\varphi & -\sin 2\varphi 	&{\bf 0}\\
				\sin 2\varphi & \cos 2\varphi	&{\bf 0}\\
				{\bf 0}	      & {\bf 0}		&{\bf 1}\\
	\end{array}
\right]\; , \;\;\;\;
f_L=	\left[
		\begin{array}{cc} 1	& {\bf 0} \\
				{\bf 0}	& g_L
		\end{array}
	\right]
\]		
We now have $f=f_R^{-1}f_{2\varphi}^{-1}f_L$.
Using the fact that
$df f^{-1}$ is of the form 
\[ 
df f^{-1}=
\left[
\begin{array}{ccccc}
	0 & * & * & * & * \\
	{}* & 0 & 0 & 0 & 0 \\
	{}* & 0 & 0 & 0 & 0 \\
	{}* & 0 & 0 & 0 & 0 \\
	{}* & 0 & 0 & 0 & 0 
\end{array}
\right]
\]
we can show that:
\begin{eqnarray}
	\mbox{tr}\left(\delta (f_R f) (f_R f)^{-1}\;
\delta(\partial_+(f_R f) (f_R f)^{-1}\right) &=& 
	\mbox{tr}\left(\delta f_R f^{-1}\; 
	\delta(\partial_+f_{R} f_{R}^{-1})\right)+\nonumber\\
&&	+\mbox{tr}\left(\delta f f^{-1}\;
	\delta(\partial_+f f^{-1})\right)+\nonumber\\
&&	+\partial_+\mbox{tr}\left(\delta f f^{-1} f_R^{-1}\delta f_R
\right)
\label{fRf}
\end{eqnarray}
On the other hand this is equal to:
\begin{eqnarray}
	\mbox{tr}\left(\delta(f_{2\varphi}^{-1}f_L)f_L^{-1}f_{2\varphi}
	\delta(\partial_+(f_{2\varphi}^{-1}f_L)f_L^{-1}f_{2\varphi})\right)
	&=&\mbox{tr}\left(f_{2\varphi}^{-1}\delta f_{2\varphi}
	\delta (f_{2\varphi}^{-1}\partial_+ f_{2\varphi})\right)+\nonumber\\
&&	+\mbox{tr}\left(\delta f_L f_L^{-1} \delta (\partial_+ f_L f_L^{-1})
\right)
\nonumber
\end{eqnarray}
This equation and Eq.(\ref{fRf}) imply that 
the symplectic structure (\ref{BoostInvariantGeometric})
is on the light cone equal to the first term in
(\ref{SymplecticFormLCSuperstring}): 
\begin{equation}
\int d\tau^+ \mbox{tr}\; \left(\delta f f^{-1} \delta(\partial_+f f^{-1})
\right) 
\end{equation}
It would be interesting to find a similar geometrical interpretation for the
symplectic structure (\ref{SymplecticFormLCSuperstring}) of the 
full superstring in $AdS_5\times S^5$.

\section*{Acknowledgments}
I would like to thank N.J.~Berkovits for discussions of the zero curvature
equations and dressing transformations, 
S.~Cherkis and A.~Tseytlin for comments, and Google Scholar for 
pointing me to the references
\cite{Bakas:1995bm} and 
\cite{Doliwa:1994bk} --- \cite{Anco:2005b}. 
This research was supported by the Sherman Fairchild 
Fellowship and in part
by the RFBR Grant No.  06-02-17383 and in part by the 
Russian Grant for the support of the scientific schools
NSh-1999.2003.2.


\providecommand{\href}[2]{#2}\begingroup\raggedright\endgroup

\end{document}